\pdfoutput=1
\documentclass[11pt,a4paper]{article}

\usepackage{arxiv}
\usepackage[utf8]{inputenc}
\usepackage[T1]{fontenc}
\usepackage{url}
\usepackage{booktabs}
\usepackage{amsfonts}
\usepackage{amsmath}
\usepackage{amssymb}
\usepackage{microtype}
\usepackage{graphicx}
\usepackage{natbib}
\usepackage{float}
\usepackage{xcolor}
\usepackage{listings}
\usepackage{tikz}
\usetikzlibrary{positioning,arrows.meta,fit,calc,backgrounds}
\usepackage{hyperref}

\tikzset{
  ev/.style={draw=black!85, line width=0.7pt, rounded corners=1.5pt, fill=white,
             align=center, font=\footnotesize, inner sep=3pt},
  de/.style={draw=black!60, line width=0.6pt, dash pattern=on 2.4pt off 1.6pt,
             rounded corners=1.5pt, fill=white, align=center, font=\footnotesize,
             inner sep=3pt},
  nt/.style={draw=black!30, line width=0.4pt, rounded corners=1.5pt,
             fill=black!4, align=center, font=\footnotesize, inner sep=3pt},
  flow/.style={-{Stealth[length=4pt]}, line width=0.55pt, black!70},
  item/.style={align=center, font=\scriptsize, inner sep=1.5pt},
  stagename/.style={align=left, anchor=west, font=\scriptsize, inner sep=1pt},
  refnote/.style={font=\scriptsize, text=black!55, inner sep=1pt},
}
\newcommand{\cmark}{\textcolor{green!45!black}{\ensuremath{\checkmark}}}
\newcommand{\xmark}{\textcolor{red!70!black}{\ensuremath{\times}}}
\newcommand{\figlegend}[1]{%
  \node[anchor=west, font=\scriptsize] at (#1) {%
    \tikz[baseline=-0.6ex]\draw[black!85, line width=0.7pt, rounded corners=1pt] (0,0) rectangle (0.55,0.22);\ evaluated (\S\ref*{sec:11})\quad
    \tikz[baseline=-0.6ex]\draw[black!60, line width=0.6pt, dash pattern=on 2.4pt off 1.6pt, rounded corners=1pt] (0,0) rectangle (0.55,0.22);\ specified or built, not evaluated\quad
    \tikz[baseline=-0.6ex]\draw[black!30, fill=black!4, line width=0.4pt, rounded corners=1pt] (0,0) rectangle (0.55,0.22);\ input or existing system};}

\graphicspath{{./}}

\title{Brain API: An Intent-Aware Control Plane for Policy-Governed Agentic Systems}

\newcommand{\authoraffiliation}{%
    Independent Researcher,\\
    Toronto, ON, Canada%
}

\author{
    \vspace{2mm} \\
    \texttt{Alexander Chernov} \\
    \vspace{1mm} \\
    \authoraffiliation
}

\renewcommand{\shorttitle}{\textit{Brain API}}

\hypersetup{
    pdftitle={Brain API: An Intent-Aware Control Plane for Policy-Governed Agentic Systems},
    pdfsubject={Computer Science, Distributed Systems, Machine Learning},
    pdfauthor={Alexander Chernov},
    pdfkeywords={Brain API, Control Plane, Intent-Aware Systems, Decision Artifacts, Policy Governance, Agentic Systems},
    colorlinks=true
}

\begin{document}

\maketitle

\begin{center}
\small
\textbf{Preprint.}
\end{center}

\vspace{0.5em}

\begin{abstract}
Contemporary cloud and distributed systems expose control through resource-centric abstractions: services, deployments, network flows, execution graphs. Agentic and tool-augmented systems have meanwhile shifted application logic toward intent-driven, adaptive execution. Existing control planes, workflow engines and service meshes lack abstractions for intent-level decision governance: they cannot represent high-level goals as first-class control objects, cannot enforce policy over the mapping from intent to execution plan, and cannot produce auditable records of \textit{why} one execution path was chosen over its alternatives. Control logic is therefore embedded in application code, leaving systems brittle, opaque and hard to govern.

We propose \textbf{Brain API}, an intent-aware control plane for policy-governed agentic systems. Its central contribution is the \textbf{decision artifact}: a durable, versioned, auditable record of how an intent became an executable plan, capturing which policies applied, which capabilities were evaluated, which alternatives were rejected, and why. A motivating use case is \textit{agentic datasets}: datasets participating as policy-governed capabilities under residency, compliance and cost constraints.

We evaluate a prototype of the decision layer against two external policy corpora we did not author. On the OPA Gatekeeper constraint library it agrees with the library's own published verdicts on 42 of 42 encodable cases, 19 admit and 23 deny. On Cedar example policies, labeled by differential testing against its reference implementation, a deliberately dissimilar domain exposed three defects in our model, including a default-allow assumption that would have inverted every authorization policy. The evaluation covers policy filtering and selection; candidate generation, context signals, ranking and plan synthesis are not measured, nor is decision latency under load.
\end{abstract}

\vspace{0.1em}
\noindent\textbf{How to cite:} Chernov, Alexander. 2026. \textit{Brain API: An Intent-Aware Control Plane for Policy-Governed Agentic Systems}. arXiv preprint arXiv:2609.21299.


\section{Introduction}
\label{sec:1}

Distributed systems have traditionally been controlled through resource-centric abstractions such as services, deployments, jobs, and tasks. Control planes in systems such as Kubernetes, service meshes, and workflow engines focus on scheduling, placement, scaling, and connectivity of these resources \citep{burns2016kubernetes,verma2015borg,schwarzkopf2013omega,hindman2011mesos}. While this model has proven effective for managing infrastructure and stateless request flows, it becomes increasingly strained in the presence of \emph{agentic} and \emph{tool-augmented} applications that operate over high-level goals, adapt their behavior at runtime, and compose heterogeneous capabilities dynamically.

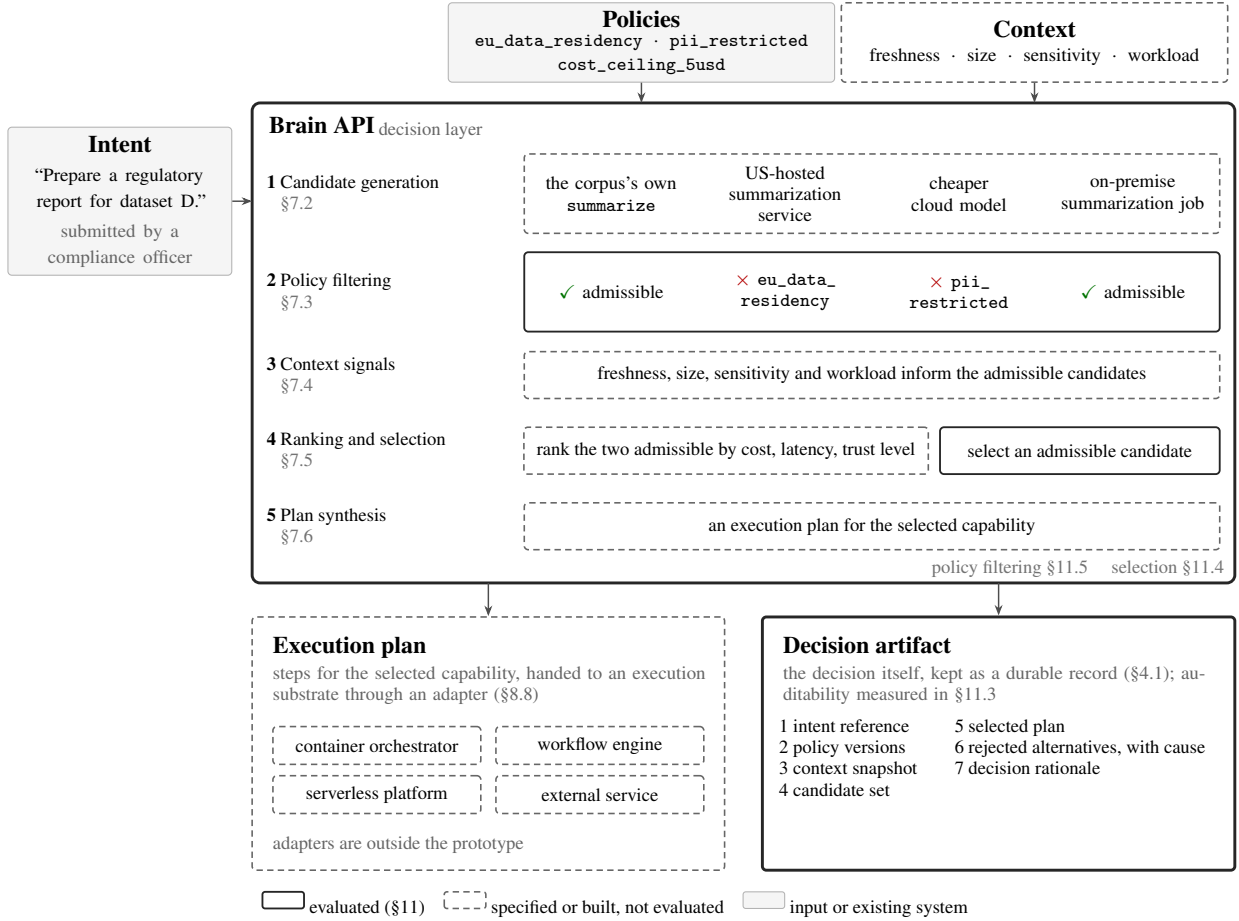
\begin{figure}[!t]
  \centering
  \begin{tikzpicture}[x=1cm, y=1cm]
  \node[nt, text width=4.9cm, minimum height=1.05cm] (pol) at (8.35,12.35)
    {\textbf{Policies}\\[-1pt]{\scriptsize\texttt{eu\_data\_residency} \,·\, \texttt{pii\_restricted}\\[-1pt]\texttt{cost\_ceiling\_5usd}}};
  \node[de, text width=4.9cm, minimum height=1.05cm] (ctx) at (13.55,12.35)
    {\textbf{Context}\\[-1pt]{\scriptsize freshness \,·\, size \,·\, sensitivity \,·\, workload}};

  \node[nt, text width=2.75cm] (intent) at (1.45,10.25)
    {\textbf{Intent}\\[1pt]{\scriptsize ``Prepare a regulatory report for dataset~D.''}\\[1pt]{\scriptsize\color{black!60} submitted by a compliance officer}};

  \draw[line width=1.1pt, rounded corners=3pt, black!85] (3.2,5.2) rectangle (16.2,11.55);
  \node[anchor=north west, font=\small\bfseries] at (3.3,11.5) {Brain API};
  \node[anchor=north west, font=\scriptsize, text=black!60] at (4.75,11.42) {decision layer};

  \node[de, minimum width=9.2cm, minimum height=1.05cm] at (11.4,10.35) {};
  \node[ev, minimum width=9.2cm, minimum height=1.05cm] at (11.4,9.05) {};
  \node[de, minimum width=9.2cm, minimum height=0.62cm] at (11.4,7.95) {};
  \node[de, minimum width=5.35cm, minimum height=0.62cm] at (9.47,6.95) {};
  \node[ev, minimum width=3.7cm, minimum height=0.62cm] at (14.15,6.95) {};
  \node[de, minimum width=9.2cm, minimum height=0.62cm] at (11.4,5.95) {};

  \node[stagename] at (3.35,10.35) {\textbf{1} Candidate generation\\\hphantom{\textbf{1} }\color{black!55}\S\ref*{sec:7.2}};
  \node[stagename] at (3.35,9.05)  {\textbf{2} Policy filtering\\\hphantom{\textbf{2} }\color{black!55}\S\ref*{sec:7.3}};
  \node[stagename] at (3.35,7.95)  {\textbf{3} Context signals\\\hphantom{\textbf{3} }\color{black!55}\S\ref*{sec:7.4}};
  \node[stagename] at (3.35,6.95)  {\textbf{4} Ranking and selection\\\hphantom{\textbf{4} }\color{black!55}\S\ref*{sec:7.5}};
  \node[stagename] at (3.35,5.95)  {\textbf{5} Plan synthesis\\\hphantom{\textbf{5} }\color{black!55}\S\ref*{sec:7.6}};

  \foreach \x/\name in {7.95/{the corpus's own\\ \texttt{summarize}},
                        10.25/{US-hosted\\ summarization service},
                        12.55/{cheaper\\ cloud model},
                        14.85/{on-premise\\ summarization job}} {
    \node[item, text width=2.15cm] at (\x,10.35) {\name};
  }
  \node[item, text width=2.15cm] at (7.95,9.05)  {\cmark\ admissible};
  \node[item, text width=2.15cm] at (10.25,9.05) {\xmark\ \texttt{eu\_data\_}\\\texttt{residency}};
  \node[item, text width=2.15cm] at (12.55,9.05) {\xmark\ \texttt{pii\_}\\\texttt{restricted}};
  \node[item, text width=2.15cm] at (14.85,9.05) {\cmark\ admissible};
  \node[item] at (11.4,7.95) {freshness, size, sensitivity and workload inform the admissible candidates};
  \node[item] at (9.47,6.95) {rank the two admissible by cost, latency, trust level};
  \node[item] at (14.15,6.95) {select an admissible candidate};
  \node[item] at (11.4,5.95) {an execution plan for the selected capability};
  \node[refnote, anchor=south east] at (16.1,5.24) {policy filtering \S\ref*{sec:11.5} \quad selection \S\ref*{sec:11.4}};

  \draw[flow] (pol.south) -- (8.35,11.55);
  \draw[flow] (ctx.south) -- (13.55,11.55);
  \draw[flow] (intent.east) -- (3.2,10.25);

  \node[de, minimum width=6.25cm, minimum height=3.35cm] (plan) at (6.325,3.075) {};
  \node[ev, line width=1.1pt, minimum width=6.25cm, minimum height=3.35cm] (art) at (13.075,3.075) {};
  \draw[flow] (6.325,5.2) -- (plan.north);
  \draw[flow] (13.075,5.2) -- (art.north);

  \node[anchor=north west, font=\footnotesize\bfseries] at (3.35,4.62) {Execution plan};
  \node[anchor=north west, font=\scriptsize, text=black!60, text width=5.9cm, align=left] at (3.35,4.22)
    {steps for the selected capability, handed to an execution substrate through an adapter (\S\ref*{sec:8.8})};
  \foreach \x/\y/\t in {4.85/3.05/container orchestrator, 7.8/3.05/workflow engine,
                        4.85/2.4/serverless platform, 7.8/2.4/external service} {
    \node[de, font=\scriptsize, minimum width=2.75cm, minimum height=0.5cm, inner sep=1.5pt] at (\x,\y) {\t};
  }
  \node[anchor=south west, font=\scriptsize, text=black!60] at (3.35,1.5) {adapters are outside the prototype};

  \node[anchor=north west, font=\footnotesize\bfseries] at (10.1,4.62) {Decision artifact};
  \node[anchor=north west, font=\scriptsize, text=black!60, text width=5.9cm, align=left] at (10.1,4.22)
    {the decision itself, kept as a durable record (\S\ref*{sec:4.1}); auditability measured in \S\ref*{sec:11.3}};
  \node[anchor=north west, font=\scriptsize, align=left] at (10.05,3.55)
    {\begin{tabular}{@{}l@{\qquad}l@{}}
       1\ intent reference & 5\ selected plan\\
       2\ policy versions & 6\ rejected alternatives, with cause\\
       3\ context snapshot & 7\ decision rationale\\
       4\ candidate set &
     \end{tabular}};

  \figlegend{3.2,0.95}
\end{tikzpicture}
  \caption{From intent to governed execution, for the running example of \S\ref{sec:1} as \S\ref{sec:10.1} walks it through the pipeline of \S\ref{sec:7}. An intent does not map directly to execution: it enters a decision in which candidates are generated, policy removes the inadmissible ones, the remainder are ranked and one is selected, and the choice is kept as a decision artifact beside the execution plan. Candidates and eliminations are those \S\ref{sec:10.1} names; it does not say which admissible candidate wins, and neither does the figure. Borders mark whether a stage was evaluated in \S\ref{sec:11} --- on the Gatekeeper corpus, not on this illustrative example: policy filtering, selection and the artifact were; candidate generation, context signals, ranking and plan synthesis were not. The artifact lists the contents defined in \S\ref{sec:4.1}; Figure~\ref{fig:anatomy} shows which of them the prototype records.}
  \label{fig:governed-decision}
\end{figure}

\textbf{Running example.} To ground the discussion, we use a concrete scenario throughout the paper. A compliance officer submits the intent: \emph{“Prepare a regulatory report for dataset D.”} Fulfilling this intent requires selecting among candidate capabilities (the dataset's own summarization, a US-hosted summarization service, a cheaper cloud model, an on-premise summarization job), enforcing governance policies (data residency constraints, access class requirements, cost ceilings), and reusing prior results only if they are sufficiently fresh and produced under the same policy regime. The system must record \emph{why} a particular execution path was chosen---not just what executed---so that the decision can be audited independently of its outcome. We use this scenario to illustrate each layer of Brain API as it is introduced, and revisit it in full in the use cases (\S\ref{sec:10.1}). Figure~\ref{fig:governed-decision} traces it through the decision layer.

Recent systems that integrate large language models, planners, and tool execution frameworks exemplify the broader shift toward intent-driven execution \citep{yao2023react,schick2023toolformer}. Instead of executing a fixed pipeline or a statically defined workflow, such systems reason over \emph{intent}---as in our running example---and decompose it into a sequence of actions that may involve services, tools, and other agents. In practice, however, the control logic governing this process is embedded inside application code or orchestration scripts. Policies related to safety, compliance, cost, or data locality are enforced in an ad hoc manner, if at all. Observability is similarly limited: operators can inspect resource metrics and logs, but rarely the \emph{decisions} that led to a particular execution path.

This situation creates three systemic problems. First, intent-level control is fragmented across applications, making it difficult to enforce global policies or reason about system-wide behavior. Second, decision-making logic is opaque and tightly coupled to execution, hindering auditing, debugging, and governance. Third, the lack of a shared control-plane abstraction prevents reuse and composition across different agentic systems and execution backends.

We argue that these problems stem from a missing abstraction: a control plane that treats \emph{intent} and \emph{policy} as first-class concerns, rather than as application-level conventions. To address this gap, we propose \textbf{Brain API}, an intent-aware control plane for policy-governed agentic systems. Brain API introduces a decision layer that sits between intent submission and execution, explicitly representing policies, capabilities, and routing logic. This layer determines \emph{what should be done} and \emph{under what constraints} before any concrete action is taken, and it produces an execution plan that can be carried out by heterogeneous backends.

\textbf{What Brain API uniquely contributes.} A natural question is whether Brain API's functionality can be assembled from existing components---for example, by combining OPA for policy evaluation, a service mesh for routing, a workflow engine for execution, and an LLM tool router for intent decomposition. We argue it cannot, for three reasons; the first two are empirical and we test them in \S\ref{sec:11}, and the third is an architectural argument that we do not claim as a measured result.

First, none of these systems produce a \textbf{decision artifact}: a first-class, versioned, auditable record that captures \emph{why} a specific execution path was chosen, which policies applied, which alternatives were considered, and under what context. Without decision artifacts, governance is retrospective at best---operators can observe what executed but not the control logic that led there. Measured over the corpus decisions, a decision artifact answers all six of our audit questions while an admission-control log answers two (\S\ref{sec:11.3}).

Second, none enforce policy at \textbf{intent level}: policies in existing systems govern individual requests, network flows, or resource states. Brain API enforces policy over the mapping from a high-level goal to a complete execution plan---before any action is taken---and propagates policy obligations through plan synthesis. We should be precise about what this does and does not distinguish: admission control also runs before execution, so pre-execution timing is \emph{not} the difference. The difference is that admission control adjudicates one proposed object, whereas intent-level enforcement selects among alternatives. Where a compliant alternative exists, Brain API reaches it in a single pass; the admission baseline rejects the proposal it was handed and, because its log names no alternative, cannot itself route to the compliant one (\S\ref{sec:11.4}). Figure~\ref{fig:adjudication} draws the distinction.

\begin{figure}[!ht]
  \centering
  \begin{tikzpicture}[x=1cm, y=1cm]
  \node[font=\small\bfseries, anchor=west] at (0,7.6) {Admission control};
  \node[nt, minimum width=3.4cm] (prop) at (3.6,6.75) {one proposed object};
  \node[ev, minimum width=3.4cm] (apol) at (3.6,5.6) {policy};
  \node[ev, minimum width=1.6cm, text=green!45!black] (adm) at (2.2,4.35) {\textbf{admit}};
  \node[ev, minimum width=3.0cm, align=center] (rej) at (5.3,4.35) {\textcolor{red!70!black}{\textbf{reject}}\\[-1pt]{\scriptsize with the violation}};
  \draw[flow] (prop) -- (apol);
  \draw[flow] (apol.south) -- ++(0,-0.3) -| (adm.north);
  \draw[flow] (apol.south) -- ++(0,-0.3) -| (rej.north);
  \node[ev, minimum width=6.4cm, align=center] (alog) at (3.6,2.95) {log: one proposal, one verdict\\[-1pt]{\scriptsize no alternative on record}};
  \draw[flow] (adm.south) -- (adm.south |- alog.north);
  \draw[flow] (rej.south) -- (rej.south |- alog.north);
  \node[font=\footnotesize, align=center, text width=6.6cm] at (3.6,1.3)
    {One proposed object is adjudicated.\\[2pt]
     {\scriptsize answers \textbf{2 of 6} audit questions (\S\ref*{sec:11.3})}};

  \draw[black!25, line width=0.5pt] (7.75,0.7) -- (7.75,7.85);

  \node[font=\small\bfseries, anchor=west] at (8.2,7.6) {Brain API};
  \node[nt, minimum width=3.4cm] (goal) at (12.1,6.85) {one intent};
  \foreach \x/\n in {10.6/A, 13.6/B} {
    \node[de, minimum width=2.2cm] (c\n) at (\x,5.95) {plan \n};
    \draw[flow] (goal.south) -- ++(0,-0.2) -| (c\n.north);
  }
  \node[ev, minimum width=5.4cm, minimum height=0.55cm] (filt) at (12.1,5.05) {};
  \node[item] at (10.6,5.05) {\xmark\ policy};
  \node[item] at (13.6,5.05) {\cmark\ compliant};
  \node[refnote, anchor=east] at (9.3,5.05) {filter};
  \foreach \n in {A,B} { \draw[flow] (c\n.south) -- (c\n.south |- filt.north); }
  \node[ev, minimum width=2.6cm] (sel) at (12.1,3.75) {\textbf{select} B};
  \draw[flow] (13.6,4.77) |- (sel.east);
  \node[de, minimum width=2.6cm] (exec) at (10.6,2.55) {execution plan};
  \node[ev, line width=1.0pt, minimum width=2.6cm] (dart) at (13.6,2.55) {decision artifact};
  \draw[flow] (sel.south) -- ++(0,-0.18) -| (exec.north);
  \draw[flow] (sel.south) -- ++(0,-0.18) -| (dart.north);
  \node[font=\footnotesize, align=center, text width=7.2cm] at (12.1,1.3)
    {One goal is resolved among alternatives,\\ and the choice is kept.\\[2pt]
     {\scriptsize answers \textbf{6 of 6} (\S\ref*{sec:11.3}); the compliant alternative is selected in \textbf{20 of 20} trials, in one pass (\S\ref*{sec:11.4})}};

  \figlegend{0,0.2}
\end{tikzpicture}
  \caption{Adjudication against selection. Both act before execution; the difference is that admission control adjudicates one proposed object, while Brain API resolves one goal among alternatives and keeps the choice. The figures under each panel are the measured consequences: the questions an auditor can answer from each record (\S\ref{sec:11.3}), and selection of the compliant alternative in one pass (\S\ref{sec:11.4}).}
  \label{fig:adjudication}
\end{figure}

Third, there is no \textbf{unified control-plane abstraction for agentic systems}: each existing system addresses one layer (infrastructure, network, workflow, or tool invocation) and exposes a different model. Brain API provides a single abstraction layered across all of them, making intent, policy, and decision-making reusable and independently evolvable across heterogeneous backends. This is a claim about system structure and modularity, and we argue rather than measure it.

\textbf{Scope.} We define the model and interfaces precisely and evaluate a prototype implementation of the decision layer against external policy corpora. The evaluation uses two independent policy corpora: the OPA Gatekeeper constraint library, 49 production Kubernetes admission policies whose test suites publish for each sample object whether it should be admitted; and the Cedar policy language's published example policies, labeled by differential testing against its reference implementation. We author the encoding of each policy into our rule grammar; we author neither the objects nor the verdicts, so an incorrect encoding manifests as disagreement with the corpus rather than as a favorable result.

Two limits on that scope should be stated at the outset. First, the corpus exercises the policy-filtering and selection stages; the context-signal and ranking stages (\S\ref{sec:7.4}--\S\ref{sec:7.5}) are \emph{not} evaluated here, because pass/fail admission decisions provide no ranking ground truth; candidate generation and plan synthesis run in every experiment but are not measured. Second, one of the three arguments we make in this section against assembling Brain API from existing components is architectural rather than empirical, and we mark it as such rather than counting it among the results.

\textbf{Contributions.} The contributions of this paper are as follows:

\begin{enumerate}
\item \textbf{Problem formulation}: We articulate the limitations of resource-centric control planes for agentic and intent-driven systems and formulate the need for an intent-aware, policy-governed control layer.
\item \textbf{System model}: We present a system model that elevates intent, policy, capability, and decision to first-class control-plane primitives.
\item \textbf{Architecture}: We describe the architecture of Brain API, including its separation of decision-making from execution and its integration with existing execution environments.
\item \textbf{Control interfaces and semantics}: We define the core control-plane interfaces and decision semantics that enable semantic routing, governed execution, and decision-level observability.
\item \textbf{Illustrative scenarios}: We present representative use cases that demonstrate how Brain API simplifies the construction and governance of adaptive, multi-step, agentic systems.
\item \textbf{An expressiveness result obtained from a production policy corpus}: We measure which real governance policies the decision layer's rule grammar can express. A rule language of scalar field/operator/value comparisons---the form the model was first specified with---covers 3 of 49 Gatekeeper policies, because real policies quantify over collections ("every container's image must come from an allowed repository"). Adding bounded quantification raises coverage to 28 of 49; aggregation and general regular expressions remain out of reach and we report them as such (\S\ref{sec:11.2}).
\item \textbf{Empirical evaluation of the two testable claims above}: conformance of the policy stage against the corpus's published verdicts (42 of 42 encodable cases, on a near-balanced allow/deny split); auditability of the decision artifact against an admission-control log (six of six audit questions versus two, with every decision re-derivable from its artifact alone); and selection among alternatives (\S\ref{sec:11.3}--\S\ref{sec:11.4}).
\end{enumerate}

\section{Background and Problem Statement}
\label{sec:2}

Control planes in modern distributed systems are primarily designed around managing resources and request flows. Systems such as Kubernetes provide declarative mechanisms for describing desired resource states, while service meshes and API gateways focus on traffic routing, resilience, and security at the level of network requests \citep{burns2016kubernetes,istio,envoy,linkerd}. Workflow engines and schedulers extend this model to multi-step computations, but they typically rely on statically defined graphs or imperative orchestration logic \citep{deelman2005pegasus,argoworkflows,temporal}.

In parallel, agentic and tool-augmented systems have introduced a different execution paradigm. Instead of executing a fixed plan, these systems operate over high-level goals and iteratively decide which actions to take based on intermediate results, context, and external constraints. The execution path is not fully known in advance and may involve heterogeneous capabilities, including microservices, data processing jobs, external APIs, and human-in-the-loop steps. While this approach increases flexibility, it also exposes a gap between \emph{what the system is trying to achieve} and \emph{how the underlying infrastructure is controlled}.

Today, this gap is typically bridged in one of two ways. In the first approach, developers encode decision logic directly into application code, often intertwined with tool invocation and error handling. Policies related to cost, data governance, or safety are implemented as conditional checks scattered throughout the codebase. In the second approach, developers rely on workflow engines or orchestration frameworks, but these systems still require explicit enumeration of steps and transitions, pushing adaptive behavior back into application logic or external controllers.

Both approaches suffer from common limitations. Decision-making is tightly coupled to execution, making it difficult to inspect, audit, or evolve independently. Policies are enforced implicitly and locally, rather than as global, declarative constraints. Observability focuses on resource usage and request metrics, but provides little insight into \emph{why} a particular execution path was chosen. As a result, agentic systems become difficult to govern, especially in regulated or cost-sensitive environments.

At the same time, existing control planes lack primitives for representing intent, policy, or semantic suitability of capabilities. Routing decisions are typically based on static rules, labels, or low-level metrics, rather than on high-level goals and constraints. Memory and context, when used, are treated as application-specific concerns rather than as inputs to control decisions. This forces each agentic system to reinvent its own control logic, leading to fragmentation and duplicated effort.

The core problem we address in this paper is therefore the absence of a unifying control-plane abstraction for intent-driven, policy-governed execution. We seek a model in which intent submission, policy enforcement, capability selection, and execution planning are explicit, inspectable, and reusable across systems and environments. Such a model should integrate with existing distributed execution substrates while providing a higher-level decision layer that makes agentic behavior governable and observable.

One motivating class of applications for this model is \emph{agentic datasets}, where datasets are no longer treated as passive storage but as policy-governed participants in multi-step reasoning pipelines---for example, selecting processing strategies, enforcing data residency and compliance constraints, or reusing prior results under freshness and similarity requirements. In such systems, datasets expose capabilities (e.g., query, transform, summarize, validate), and their participation in an execution is mediated by the same intent, policy, and decision mechanisms described in this paper. While agentic datasets provide a concrete and compelling use case, the Brain API model itself is intentionally general and applies to any domain in which high-level intent must be mapped to governed execution across heterogeneous capabilities.

\section{Design Goals and Non-Goals}
\label{sec:3}

Brain API is designed to serve as a control plane for intent-driven and agentic systems operating in heterogeneous distributed environments. Our design is guided by the following goals.

\textbf{Intent-first control.} The primary unit of interaction with the control plane is an \emph{intent}, representing a high-level goal rather than a specific sequence of actions. The system should reason over intents and derive concrete execution plans from them.

\textbf{Policy as a first-class primitive.} Policies governing safety, compliance, cost, data locality, or operational constraints must be explicit and declarative. The control plane should enforce these policies during decision-making rather than relying on ad hoc checks in application code \citep{opa,abadi1993logic,detreville2002binder}.

\textbf{Separation of decision and execution.} Decision-making about what to do and where to do it should be logically separated from the act of execution. This separation enables independent evolution of policies, routing logic, and execution backends, and it supports auditing and inspection of decisions.

\textbf{Heterogeneous capability integration.} The system must support a diverse set of capabilities, including services, batch jobs, tools, and other agents, without imposing a single execution model. Capabilities should be discoverable and selectable based on semantic and policy-relevant attributes.

\textbf{Decision-level observability.} Operators should be able to observe and audit not only resource usage and execution outcomes, but also the decisions that led to a particular execution path, including which policies were applied and which alternatives were considered \citep{sigelman2010dapper,buneman2001why,cheney2009provenance}.

\textbf{Pluggable context and memory.} While context and memory may influence routing and planning decisions, the control plane should not mandate a specific storage or representation. Instead, it should provide hooks for integrating different context and memory providers.

We also explicitly state several non-goals. Brain API is \textbf{not} intended to:

\begin{itemize}
\item \textbf{Replace existing execution engines, workflow systems, or orchestration platforms.} It complements them by providing a higher-level decision layer. Existing backends are treated as interchangeable executors.
\item \textbf{Function as a real-time, low-latency scheduler.} Introducing a decision layer adds indirection. Workloads with sub-millisecond routing requirements are better served by data-plane mechanisms such as service meshes or load balancers.
\item \textbf{Prescribe a specific policy language or engine.} The architecture is compatible with OPA Rego, Cedar, Datalog-based systems, or custom rule engines. The choice is left to the deploying organization.
\item \textbf{Provide policy verification or formal analysis.} Brain API enforces policies at runtime but does not statically verify that a policy set is consistent, complete, or free of conflicts. Policy authoring tooling is a complementary concern.
\item \textbf{Serve as a human-in-the-loop workflow engine.} While policies may insert approval steps into an execution plan, Brain API does not itself implement human task management or case-management interfaces.
\item \textbf{Implement planners or large language models.} These are pluggable components that may back the decision engine; Brain API defines the control-plane contract around them, not their internals.
\end{itemize}

\section{System Model}
\label{sec:4}

In this section, we present the system model underlying Brain API. The model introduces a set of first-class abstractions---\emph{Intent}, \emph{Policy}, \emph{Capability}, \emph{Context}, \emph{Decision}, and \emph{Execution Plan}---that together define how high-level goals are transformed into governed, observable actions in a distributed environment. The purpose of this model is not to prescribe a specific implementation, but to provide a precise conceptual framework that separates decision-making from execution while remaining compatible with existing distributed systems substrates.

\subsection{Core Entities}
\label{sec:4.1}

We define the following core entities.

\textbf{Intent.} An intent represents a high-level goal or objective submitted to the control plane. Unlike a request in a traditional API, an intent does not specify a concrete sequence of actions. Instead, it captures \emph{what} the user or system wishes to achieve, possibly accompanied by parameters, constraints, or preferences. Examples include “analyze this dataset,” “generate a compliance report,” or “triage this incident.” Intents are treated as declarative statements of desired outcomes.

\textbf{Policy.} A policy is a declarative constraint or rule that governs how intents may be realized. Policies may encode requirements related to safety, compliance, cost, data locality, security, performance, or organizational governance. In Brain API, policies are not embedded in application logic; they are explicit inputs to the decision process and are evaluated whenever an intent is mapped to concrete actions.

\textbf{Capability.} A capability represents an executable unit that can contribute to satisfying an intent. Capabilities may correspond to microservices, batch jobs, tools, external APIs, or other agents. Each capability is described by metadata that characterizes its semantics, requirements, and operational properties, such as input and output types, cost model, trust level, data access constraints, or performance characteristics.

\textbf{Context.} Context captures dynamic information relevant to decision-making, including runtime state, environmental conditions, user attributes, historical execution traces, or external signals. Context is distinct from policy in that it describes \emph{what is currently true}, rather than \emph{what must always hold}. Context may influence which capabilities are suitable for a given intent at a particular moment.

\textbf{Decision.} A decision is the outcome of applying policies and context to an intent and a set of available capabilities. It represents a resolved choice about \emph{how} an intent will be realized, including which capabilities will be used, in what order, and under which constraints. Decisions are first-class artifacts in Brain API and are subject to inspection, auditing, and observability.

\textbf{Definition (Decision Artifact).} A \emph{decision artifact} is a durable control-plane object that records the complete reasoning process used to transform an intent into an executable plan. A decision artifact contains:

\begin{enumerate}
\item \emph{Intent reference} --- the submitted intent, including its parameters and preferences.
\item \emph{Policy versions} --- the set of policies evaluated, identified by version, so that policy evolution can be tracked.
\item \emph{Context snapshot} --- the context signals that were active at decision time (e.g., region, cost envelope, sensitivity level).
\item \emph{Candidate set} --- all capabilities considered, before and after policy filtering.
\item \emph{Selected plan} --- the execution plan that was chosen, including the sequence and structure of steps.
\item \emph{Rejected alternatives} --- capabilities or plans that were eliminated, each annotated with the policy or criterion that caused rejection.
\item \emph{Decision rationale} --- scoring or ranking evidence that explains why the selected plan was preferred over remaining alternatives.
\end{enumerate}

Decision artifacts are persistent objects in the control plane, not transient log entries. They can be retrieved, replayed under a modified policy, and compared across time. A concrete schema is given in \S\ref{sec:6.8}. This definition distinguishes Brain API from traditional control planes, which record \emph{events} (what happened) but not \emph{decisions} (why it happened that way). The consequences are substantial: decision artifacts enable post-hoc compliance auditing, reproducible reasoning, forensic debugging, and governance-driven policy refinement in ways that event logs alone cannot support.

\textbf{Execution Plan.} An execution plan is a concrete, executable representation derived from a decision. It specifies the sequence or structure of actions to be carried out by one or more execution backends. While a decision captures the rationale and constraints, the execution plan captures the operational steps that will be performed.

\subsection{Decision Function}
\label{sec:4.2}

At the heart of the model is a decision function that maps intents to execution plans under policy and context constraints. Conceptually, this can be expressed as:

\begin{center}
\fbox{\parbox{0.9\textwidth}{
Decision = F(Intent, Policies, Context, Capabilities)
}}
\end{center}

where the function \emph{F} produces both a decision artifact and an associated execution plan. This function is not assumed to be purely static or purely deterministic; it may incorporate ranking, filtering, or optimization over available capabilities. However, the outcome of the function is always an explicit decision that can be inspected and reasoned about.

This formulation makes several properties explicit. First, policy enforcement is an integral part of decision-making rather than a post-hoc validation step. Second, context is treated as a first-class input that may influence routing, selection, or decomposition of actions. Third, capabilities are not invoked directly by the intent submitter, but are selected by the control plane based on semantic and policy-relevant attributes.

\subsection{Separation of Concerns}
\label{sec:4.3}

The system model enforces a clear separation between three concerns:

\begin{enumerate}
\item \textbf{Intent specification}
\begin{itemize}
\item Captures the desired outcome without committing to an execution strategy.
\end{itemize}
\item \textbf{Decision-making}
\begin{itemize}
\item Resolves the intent into a governed and context-aware plan using policies and capability metadata.
\end{itemize}
\item \textbf{Execution}
\begin{itemize}
\item Carries out the plan using concrete backends and runtime systems.
\end{itemize}
\end{enumerate}

This separation allows policies and routing logic to evolve independently of execution mechanisms, and it enables the system to expose decision-level observability without entangling it with low-level operational details.

\subsection{Determinism, Adaptivity, and Re-evaluation}
\label{sec:4.4}

The model permits both deterministic and adaptive behavior. For a fixed set of inputs---intent, policies, context, and capabilities---the decision function may be deterministic, yielding reproducible decisions that support auditability and debugging. At the same time, changes in context or policy may trigger re-evaluation, leading to different decisions for the same intent at different times.

Importantly, re-evaluation is an explicit and observable process: when a decision is revised due to changing context or policy, the system produces a new decision artifact and, if necessary, a new execution plan. This makes adaptation a controlled and inspectable aspect of the control plane rather than an implicit side effect of application logic.

\subsection{Relationship to Distributed Execution Substrates}
\label{sec:4.5}

The system model does not assume a specific execution substrate. Execution plans may target container orchestrators, workflow engines, serverless platforms, or external services \citep{burns2016kubernetes,verma2015borg,schwarzkopf2013omega,deelman2005pegasus}. From the perspective of the model, these substrates are interchangeable backends that consume execution plans and produce outcomes. This abstraction boundary allows Brain API to integrate with existing distributed systems while introducing a higher-level, intent- and policy-aware decision layer above them.

In summary, the system model establishes intent, policy, capability, context, and decision as explicit control-plane concepts. By making the transformation from intent to execution both governed and observable, it provides the conceptual foundation for the architecture and interfaces described in the following sections.

\subsection{Decision Artifacts vs. Prior Control-Plane Primitives}
\label{sec:4.6}

A central conceptual contribution of Brain API is elevating \emph{decisions} to the same status that resources, tasks, and deployments occupy in traditional control planes. Table~\ref{tab:primitives} contrasts what existing systems store as their primary control-plane objects with what Brain API introduces.

\begin{table}[ht]
\centering
\caption{Primary control-plane objects in existing systems vs.\ Brain API. Among the systems compared here, only Brain API records the complete reasoning process—not just outcomes or states—as a durable, first-class object.}
\label{tab:primitives}
\resizebox{\textwidth}{!}{%
\begin{tabular}{lll}
\toprule
\textbf{System} & \textbf{Primary control-plane object} & \textbf{What it records} \\
\midrule
Kubernetes & Resource (Pod, Deployment, Job) & Desired and actual state \\
Workflow engine (Argo, Temporal) & Task / activity state & Execution progress and outcomes \\
Policy engine (OPA) & Rule evaluation result & Pass / deny for a single request \\
Service mesh (Istio, Linkerd) & Traffic policy & Network-level routing rules \\
\textbf{Brain API} & \textbf{Decision artifact} & \textbf{Why a plan was chosen: intent, policies, context, candidates, rationale} \\
\bottomrule
\end{tabular}%
}
\end{table}

The distinction between recording an \emph{event} ("summarizer\_v2 was invoked at 10:15") and recording a \emph{decision} ("summarizer\_v2 was selected because summarizer\_v1 failed the trust-level policy and the cloud alternative failed the data-residency policy, given the EU-west context and a medium cost ceiling") is what enables the governance, reproducibility, and auditability properties we claim. Event logs tell you what happened; decision artifacts tell you why, and allow the why to be independently audited, replayed, and contested.

\section{Architecture Overview}
\label{sec:5}

\begin{figure}[!t]
  \centering
  \begin{tikzpicture}[x=1cm, y=1cm]
  \node[nt, minimum width=7.2cm] (users) at (7.2,10.25) {users and agentic systems};
  \draw[flow] (users.south) -- node[right, refnote] {intents} (7.2,9.5);

  \draw[line width=1.1pt, rounded corners=3pt, black!85] (0.2,4.75) rectangle (16.2,9.5);
  \node[anchor=north west, font=\small\bfseries] at (0.3,9.45) {Brain API};
  \node[anchor=north west, font=\scriptsize, text=black!60] at (1.75,9.37) {control plane};

  \node[de, minimum width=3.6cm] (ingress) at (7.2,8.8) {Intent Ingress};
  \node[ev, minimum width=3.6cm, minimum height=0.95cm, line width=0.9pt] (engine) at (7.2,7.35)
    {\textbf{Decision Engine}\\[-1pt]{\scriptsize selects among alternatives \S\ref*{sec:11.4}}};
  \node[ev, minimum width=3.3cm] (policy) at (2.55,7.35) {Policy Engine\\[-1pt]{\scriptsize\S\ref*{sec:11.5}}};
  \node[de, minimum width=3.3cm] (caps) at (11.6,7.95) {Capability Registry};
  \node[de, minimum width=3.3cm] (ctx) at (11.6,6.75) {Context Providers};
  \draw[flow] (ingress) -- (engine);
  \draw[flow] (policy) -- (engine);
  \draw[flow] (caps.west) -- ++(-0.45,0) |- ($(engine.east)+(0,0.2)$);
  \draw[flow] (ctx.west) -- ++(-0.45,0) |- ($(engine.east)+(0,-0.2)$);

  \node[ev, line width=1.1pt, minimum width=3.5cm] (art) at (5.25,5.45) {Decision artifact\\[-1pt]{\scriptsize\S\ref*{sec:11.3}}};
  \node[de, minimum width=3.5cm] (plan) at (9.15,5.45) {Execution plan};
  \draw[flow] (engine.south) -- ++(0,-0.25) -| (art.north);
  \draw[flow] (engine.south) -- ++(0,-0.25) -| (plan.north);

  \node[de, text width=1.85cm, minimum height=4.1cm] (obs) at (14.9,7.1)
    {Observability Plane\\[3pt]{\scriptsize records intents, policies applied, alternatives considered, plans and outcomes}};

  \draw[black!40, line width=0.5pt, dash pattern=on 1pt off 1.5pt] (0.2,4.3) -- (16.2,4.3);
  \node[refnote, anchor=south west] at (0.25,4.32) {control plane};
  \node[refnote, anchor=north west] at (0.25,4.28) {data plane};

  \node[de, minimum width=11.2cm] (adapters) at (7.2,3.55) {Execution Adapters \quad{\scriptsize(\S\ref*{sec:8.8}; outside the prototype)}};
  \draw[flow] (plan.south) -- (plan.south |- adapters.north);
  \foreach \x/\t [count=\i] in {2.1/container orchestrators, 5.5/workflow engines, 8.9/serverless platforms, 12.3/external services} {
    \node[nt, minimum width=3.15cm, font=\scriptsize] (sub\i) at (\x,2.35) {\t};
    \draw[flow] (\x,3.55 |- adapters.south) -- (sub\i.north);
  }
  \node[font=\scriptsize, text=black!65, anchor=west] at (0.2,1.55)
    {existing execution substrates, unchanged: Brain API sits above them, not in place of them (\S\ref*{sec:5})};

  \figlegend{0.2,0.85}
\end{tikzpicture}
  \caption{Where Brain API sits. The components of \S\ref{sec:5.2} form a control plane above the execution substrates, which are used unchanged through execution adapters; the control/data plane boundary is that of \S\ref{sec:5.3}. Solid components are those \S\ref{sec:11} measured: the policy engine (\S\ref{sec:11.5}), selection in the decision engine (\S\ref{sec:11.4}) and the decision artifact (\S\ref{sec:11.3}). The prototype is the decision layer only.}
  \label{fig:stack}
\end{figure}
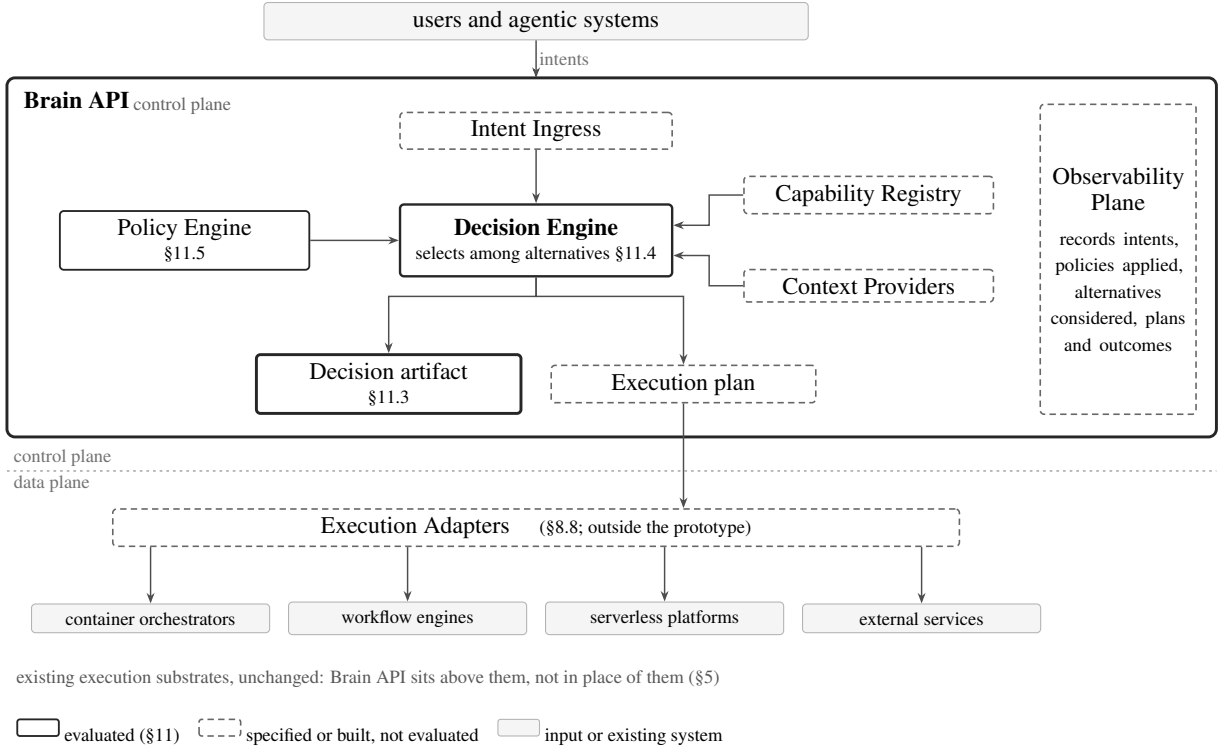

In this section, we present the high-level architecture of Brain API and describe how its components realize the system model introduced in the previous section. The architecture follows a control-plane / data-plane separation, in which intent interpretation and decision-making are centralized in a control plane, while concrete actions are carried out by heterogeneous execution backends \citep{burns2016kubernetes,verma2015borg,schwarzkopf2013omega}. This separation allows Brain API to introduce intent- and policy-aware control without replacing existing distributed execution substrates (Figure~\ref{fig:stack}).

\subsection{Architectural Principles}
\label{sec:5.1}

The architecture is guided by three principles. First, \textbf{decision-making is explicit and inspectable}: the system produces first-class decision artifacts before any execution occurs. Second, \textbf{policy enforcement is centralized} in the control plane rather than scattered across applications and tools \citep{opa,abadi1993logic,detreville2002binder}. Third, \textbf{execution is delegated} to existing systems, which are treated as interchangeable backends consuming execution plans.

These principles ensure that Brain API can be integrated incrementally into existing environments while providing a coherent control layer for agentic and intent-driven workloads.

\subsection{Component Overview}
\label{sec:5.2}

At a high level, the architecture consists of the following components:

\textbf{Intent Ingress.} The intent ingress provides the external interface through which users or systems submit intents to the control plane. It is responsible for validating intent schemas, authenticating callers, and normalizing requests into a canonical internal representation.

\textbf{Policy Engine.} The policy engine stores and evaluates declarative policies. Given an intent and relevant context, it determines which constraints apply and how they restrict the set of admissible execution strategies. The policy engine does not execute actions; it produces constraints and decisions that shape routing and planning.

\textbf{Capability Registry.} The capability registry maintains metadata about available capabilities, including their semantic descriptions and operational properties. This registry enables the control plane to reason about which capabilities are suitable candidates for satisfying a given intent under current policies and context.

\textbf{Decision Engine.} The decision engine is the core of the control plane. It combines the intent, applicable policies, current context, and available capabilities to produce a decision artifact and an associated execution plan. This component embodies the decision function defined in the system model and is responsible for selection, composition, and ordering of capabilities.

\textbf{Execution Adapters.} Execution adapters translate execution plans into concrete actions on specific backends, such as container orchestrators, workflow engines, serverless platforms, or external services \citep{deelman2005pegasus,argoworkflows,temporal}. Each adapter implements the interface required to submit, monitor, and, if necessary, cancel or modify executions on its target substrate.

\textbf{Observability Plane.} The observability plane collects and exposes information about decisions and executions. It records which intents were submitted, which policies were applied, which alternatives were considered, and which execution plans were produced, as well as the outcomes of their execution \citep{sigelman2010dapper,buneman2001why,cheney2009provenance}. This plane enables auditing, debugging, and governance at the level of decisions rather than only at the level of resources or requests.

\textbf{Context Providers.} Context providers supply dynamic information used during decision-making, such as runtime state, environmental signals, historical traces, or external annotations. They are pluggable components that allow the control plane to incorporate domain-specific knowledge without hard-coding it into the decision logic.

\begin{figure}[!ht]
  \centering
  \includegraphics[width=0.95\textwidth]{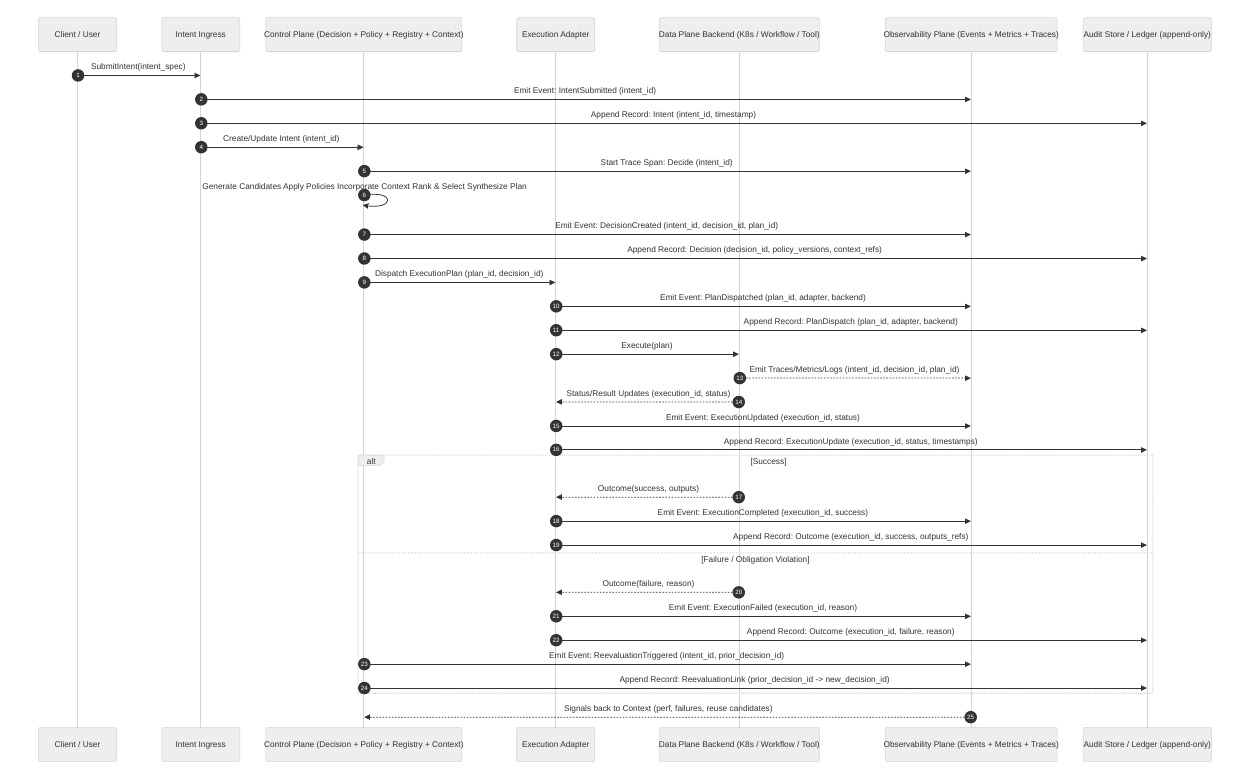}
  \caption{Decision-level observability and audit trail across control and data planes. Brain API emits correlated events and traces for intent submission, decision creation, plan dispatch, and execution outcomes using stable identifiers (intent, decision, plan, execution). Decision inputs (policy versions and context references) and outputs (decision and execution plan artifacts) are persisted as append-only audit records, enabling post hoc reconstruction of why an execution path was chosen and under which constraints. Numbered markers indicate the chronological sequence of control- and data-plane events from intent submission through execution, outcome recording, and possible re-evaluation. This is a design view: none of its components is measured as drawn; Figure~\ref{fig:stack} marks which ones \S\ref{sec:11} evaluated.}
  \label{fig:observability}
\end{figure}

Figure~\ref{fig:observability} illustrates the end-to-end control and execution lifecycle. The numbered steps indicate the chronological sequence: (1) intent submission, (2--4) intent registration, (5--8) decision synthesis, (9--11) plan dispatch, (12--16) execution and status reporting, (17--22) outcome recording for both the success and failure paths, and (23--25) re-evaluation and the signals fed back to context. This sequence highlights how decision artifacts and execution telemetry are correlated and persisted for auditability.

\subsection{Control Plane and Data Plane}
\label{sec:5.3}

Brain API enforces a clear separation between the control plane and the data plane. The control plane comprises the intent ingress, policy engine, capability registry, decision engine, and observability plane. Its responsibility is to interpret intents, apply policies, select and compose capabilities, and produce execution plans and decision artifacts.

The data plane consists of the execution backends and the execution adapters that interface with them. Its responsibility is to carry out the actions specified in execution plans and to report outcomes and status back to the control plane. This separation allows the control plane to remain agnostic to the details of execution environments while still exerting governance and providing observability over the system’s behavior.

\subsection{Execution Flow}
\label{sec:5.4}

A typical execution proceeds as follows. An intent is submitted through the intent ingress and normalized. The decision engine queries the capability registry to obtain candidate capabilities and consults context providers to gather relevant dynamic information. The policy engine evaluates applicable policies and produces constraints. The decision engine then applies the decision function to derive a decision artifact and an execution plan that satisfies the policies and is appropriate for the current context. The execution plan is handed off to the appropriate execution adapters, which submit it to the chosen backends. Throughout this process, the observability plane records the intent, the applied policies, the decision, and the execution outcomes.

\subsection{Integration with Existing Systems}
\label{sec:5.5}

The architecture is designed to integrate with existing distributed systems rather than to replace them. Brain API can sit alongside container orchestrators, workflow engines, and service meshes, providing an additional control layer that governs how these systems are used in response to high-level intents. By treating execution substrates as backends and exposing decisions as first-class artifacts, the architecture enables incremental adoption and coexistence with established infrastructure.

In summary, the architecture of Brain API realizes the system model by introducing an explicit, policy-aware decision layer between intent submission and execution. This layer provides a unifying control plane for agentic systems while leveraging existing data-plane technologies for scalable and reliable execution.

\section{Brain API: Control Plane Interfaces}
\label{sec:6}

In this section, we describe the core control-plane interfaces exposed by Brain API. These interfaces define how intents, policies, and capabilities are registered and managed, how decisions are produced and inspected, and how executions are observed. The goal is not to prescribe a concrete wire protocol or implementation language, but to specify the minimal set of operations required to realize the system model and architecture described in the previous sections.

\subsection{Design Principles}
\label{sec:6.1}

The interface design follows three principles:

\begin{enumerate}
\item \textbf{Intents, policies, and capabilities are first-class objects}
\begin{itemize}
\item Can be created, updated, and queried independently.
\end{itemize}
\item \textbf{Decisions are explicit artifacts}
\begin{itemize}
\item Can be retrieved and inspected before and after execution.
\end{itemize}
\item \textbf{Execution is observable and controllable}
\begin{itemize}
\item Achieved through uniform abstractions, regardless of the underlying backend.
\end{itemize}
\end{enumerate}

\subsection{Intent Interface}
\label{sec:6.2}

The intent interface allows clients to submit, query, and manage intents. Conceptually, it provides operations of the following form:

\begin{itemize}
\item \texttt{SubmitIntent(intent\_spec) -> intent\_id}
\item \texttt{GetIntent(intent\_id) -> intent\_spec}
\item \texttt{ListIntents(filter) -> [intent\_id]}
\end{itemize}

An \texttt{intent\_spec} describes the desired outcome and may include parameters, constraints, or preferences. The submission of an intent does not directly trigger execution; instead, it initiates the decision-making process in the control plane. The returned \texttt{intent\_id} serves as a stable handle for tracking the lifecycle of the intent, its associated decisions, and any resulting executions.

\subsection{Policy Interface}
\label{sec:6.3}

The policy interface manages declarative constraints that govern decision-making. It provides operations such as:

\begin{itemize}
\item \texttt{RegisterPolicy(policy\_spec) -> policy\_id}
\item \texttt{UpdatePolicy(policy\_id, policy\_spec)}
\item \texttt{GetPolicy(policy\_id) -> policy\_spec}
\item \texttt{ListPolicies(filter) -> [policy\_id]}
\end{itemize}

A \texttt{policy\_spec} encodes rules or constraints over intents, capabilities, context, or execution environments. Policies are evaluated by the policy engine during decision-making and are not embedded in application logic. This separation allows policies to be evolved, audited, and reasoned about independently of the code that submits intents or executes actions.

\subsection{Capability Interface}
\label{sec:6.4}

The capability interface exposes operations for registering and discovering executable capabilities:

\begin{itemize}
\item \texttt{RegisterCapability(capability\_spec) -> capability\_id}
\item \texttt{UpdateCapability(capability\_id, capability\_spec)}
\item \texttt{GetCapability(capability\_id) -> capability\_spec}
\item \texttt{ListCapabilities(filter) -> [capability\_id]}
\end{itemize}

A \texttt{capability\_spec} describes the semantics and operational properties of an executable unit, such as its inputs and outputs, cost model, trust level, or data access constraints. This metadata enables the decision engine to reason about suitability and compatibility when selecting capabilities for a given intent under applicable policies.

\subsection{Decision Interface}
\label{sec:6.5}

The decision interface provides visibility into the outcomes of the control plane’s decision-making process. Typical operations include:

\begin{itemize}
\item \texttt{GetDecision(intent\_id) -> decision\_artifact}
\item \texttt{ListDecisions(filter) -> [decision\_id]}
\end{itemize}

A \texttt{decision\_artifact} records the selected capabilities, the structure of the execution plan, the policies that were applied, and, optionally, alternative candidates that were considered and rejected. By exposing decisions as first-class objects, Brain API enables auditing, debugging, and governance at the level of control logic rather than only at the level of execution outcomes.

\subsection{Execution Interface}
\label{sec:6.6}

The execution interface allows clients and operators to observe and, when necessary, control the execution of plans derived from decisions. Conceptual operations include:

\begin{itemize}
\item \texttt{GetExecution(decision\_id) -> execution\_status}
\item \texttt{ListExecutions(filter) -> [execution\_id]}
\item \texttt{CancelExecution(execution\_id)}
\end{itemize}

The \texttt{execution\_status} includes information about progress, success or failure, and any intermediate results exposed by the underlying execution backend. While the specifics of execution management depend on the backend, Brain API provides a uniform abstraction for tracking and governing executions across heterogeneous systems.

\subsection{Introspection and Audit}
\label{sec:6.7}

In addition to the interfaces above, Brain API exposes introspection and audit capabilities that allow operators to query how and why particular decisions were made. This includes retrieving the set of policies that were in force at the time of decision-making, the context inputs that were considered, and the rationale for selecting one capability or plan over alternatives. These interfaces are essential for compliance, debugging, and post hoc analysis in regulated or safety-critical environments.

\subsection{Illustrative Artifact Schemas}
\label{sec:6.8}

To ground the interface specification, we provide illustrative JSON schemas for the three primary artifact types. These are representative and non-normative; implementations may use different serialization formats.

\textbf{Intent specification} (\texttt{intent\_spec}):
\begin{verbatim}
{
  "intent_id": "i-7f3a1b",
  "submitter": "user:alice",
  "goal": "produce-regulatory-summary",
  "parameters": {
    "dataset": "ds://compliance/q4-2025",
    "output_format": "pdf",
    "deadline_iso": "2026-03-05T18:00:00Z"
  },
  "preferences": {
    "max_cost_usd": 5.00
  }
}
\end{verbatim}

\textbf{Capability specification} (\texttt{capability\_spec}):
\begin{verbatim}
{
  "capability_id": "cap-summarize-v2",
  "name": "Regulatory Summarizer",
  "input_types": ["tabular", "text"],
  "output_types": ["pdf", "json"],
  "trust_level": "elevated",
  "cost_model": { "type": "per-call", "usd": 1.20 },
  "data_residency": ["eu-west-1", "ca-central-1"]
}
\end{verbatim}

\textbf{Decision artifact} (\texttt{decision\_artifact}):
\begin{verbatim}
{
  "decision_id": "d-9c21f0",
  "intent_id": "i-7f3a1b",
  "timestamp": "2026-03-04T10:15:00Z",
  "policy_versions": ["pol-data-residency@v3", "pol-cost-ceiling@v1"],
  "context_refs": ["ctx-user-profile", "ctx-cluster-load"],
  "selected_capability": "cap-summarize-v2",
  "rejected_candidates": [
    { "capability_id": "cap-summarize-v1", "reason": "trust_level insufficient" },
    { "capability_id": "cap-cloud-summarizer", "reason": "data_residency violation" }
  ],
  "execution_plan_id": "ep-f41b88"
}
\end{verbatim}

The decision artifact is the primary audit record: it links the selected plan to the policies and context that governed the choice, and preserves rejected alternatives with rejection reasons.

\subsection{Discussion}
\label{sec:6.9}

Together, these interfaces define the contract between clients, operators, and the Brain API control plane. By making intents, policies, capabilities, decisions, and executions explicit and independently addressable, the API supports a clean separation of concerns and enables governed, observable, and evolvable agentic systems.

\section{Decision and Routing Semantics}
\label{sec:7}

This section specifies the semantics by which Brain API maps intents to governed execution plans. We describe the stages of candidate generation, policy filtering, ranking and selection, and plan synthesis, and we clarify the role of context and memory in these stages \citep{yao2023react,schick2023toolformer}. The goal is to make decision-making explicit, reproducible, and inspectable while permitting adaptive behavior under changing conditions.

\subsection{Inputs and Artifacts}
\label{sec:7.1}

The decision process consumes four classes of inputs: an \textbf{intent} submitted by a client, a set of active \textbf{policies}, the current \textbf{context} (including optional memory-derived signals), and the set of registered \textbf{capabilities}. It produces two first-class artifacts: a \textbf{decision artifact} and an associated \textbf{execution plan}. The decision artifact records the rationale and constraints; the execution plan encodes operational steps.

\subsection{Candidate Generation}
\label{sec:7.2}

Given an intent, the control plane first generates a set of candidate capability compositions. This stage is intentionally permissive: it considers all capabilities whose semantic descriptors are compatible with the intent’s objective and parameters. Compatibility is determined by schema-level and semantic predicates (e.g., input/output types, declared effects, trust domains). Candidate generation may yield single-step candidates or multi-step compositions when the intent requires decomposition.

The output of this stage is a candidate set C = \{$c_1$, $c_2$, …, $c_n$\}, where each $c_i$ denotes a potential plan skeleton (possibly partial) annotated with capability metadata.

\subsection{Policy Filtering}
\label{sec:7.3}

Policies are then applied to restrict the candidate set. Each policy evaluates predicates over the intent, context, and candidate metadata. Candidates that violate any mandatory policy are eliminated. Policies may also attach obligations or conditions to remaining candidates (e.g., required isolation level, data locality constraints, approval steps), which are propagated forward into plan synthesis.

Formally, let $P$ be the set of active policies and $F_P(C)$ the filtering operator induced by $P$. The filtered set $C' = F_P(C)$ contains only candidates that satisfy all mandatory constraints, with accumulated obligations recorded for subsequent stages.

\subsection{Contextual Constraints and Memory Signals}
\label{sec:7.4}

Context and memory-derived signals influence both filtering and ranking. Contextual constraints (e.g., current load, incident state, user attributes) may eliminate candidates that are otherwise policy-compliant. Memory signals (e.g., recent outcomes, cached results, historical performance) provide additional features that inform preference but do not, by default, override mandatory policies \citep{lewis2020rag,khandelwal2020knnlm}. The model therefore treats context and memory as first-class inputs whose effects are explicit and recorded in the decision artifact.

\subsection{Ranking and Selection}
\label{sec:7.5}

From the filtered set $C'$, the decision engine computes a ranking using a scoring function $s: C' \times \Gamma \rightarrow \mathbb{R}$ that aggregates multiple criteria, such as estimated cost, latency, reliability, compliance risk, or semantic suitability. The scoring function is configurable and may incorporate context- and memory-derived features. The top-ranked candidate (or a small Pareto-optimal set, depending on configuration) is selected \citep{hwang1981mcdm,marler2004mcdm}.

Importantly, the ranking phase is advisory rather than permissive: policies define hard constraints enforced in \S\ref{sec:7.3}, while ranking expresses preferences within the admissible set. The decision artifact records both the selected candidate and a summary of alternatives and their scores to support auditability.

\subsubsection{Formal Properties of the Decision Function}
\label{sec:7.5.1}

While the specific implementation of $F$ may vary across deployments, we require the following properties to hold for any conforming implementation.

\textbf{Policy soundness.} The selected execution plan $\pi^*$ satisfies all mandatory policy constraints: for every mandatory policy $p \in P$, $\pi^* \in \mathrm{Admissible}(p)$. The system must never select a non-compliant candidate regardless of its score.

\textbf{Best-effort completeness.} If the filtered candidate set $C' = F_P(C)$ is non-empty, $F$ returns a selected plan. If $C' = \emptyset$, $F$ signals an \emph{unsatisfiable intent} rather than silently selecting a non-compliant alternative. Callers must be able to distinguish a refused intent from a failed execution.

\textbf{Monotone filtering.} Adding a policy can only restrict the admissible set: for any additional policy $p$, $F_{P \cup \{p\}}(C) \subseteq F_P(C)$. This ensures that stricter policy regimes cannot inadvertently re-admit previously excluded candidates.

\textbf{Score independence from hard constraints.} The scoring function $s$ is evaluated only over the filtered set $C'$; it cannot promote a policy-violating candidate above a compliant one.

\textbf{Reproducibility.} For a fixed input tuple $(\mathcal{I}, P, \Gamma, \mathcal{C})$ and a fixed scoring configuration $\sigma$, $F$ is deterministic. Changes to any input produce a new, distinctly versioned decision artifact.

These properties establish a minimal contract for any implementation of $F$. They are sufficient to guarantee policy safety and auditability while leaving the scoring strategy and planner implementation open to domain-specific choices.

\subsection{Plan Synthesis}
\label{sec:7.6}

The selected candidate is then transformed into a concrete execution plan. Plan synthesis resolves obligations attached during policy filtering (e.g., inserting approval steps, selecting specific regions or trust domains), binds abstract capability references to concrete endpoints or adapters, and orders steps according to declared dependencies. The result is an executable structure consumable by the appropriate execution adapter.

\subsection{Determinism and Reproducibility}
\label{sec:7.7}

For a fixed tuple (Intent, P, Context, Capabilities) and a fixed scoring configuration, the decision process can be deterministic, yielding reproducible decisions. Brain API records the inputs, policy versions, and scoring configuration used, enabling post hoc reconstruction and audit. When any of these inputs change, the system may re-evaluate and produce a new decision artifact, which is explicitly linked to the prior one.

\subsection{Failure, Fallbacks, and Re-routing}
\label{sec:7.8}

If execution fails or violates obligations at runtime, the control plane may trigger re-evaluation. In such cases, the prior decision artifact and execution outcomes become part of context, and the process repeats from candidate generation or from ranking, depending on policy. This enables controlled fallbacks and re-routing while preserving an auditable trail of decisions and outcomes.

\subsection{Summary}
\label{sec:7.9}

Decision and routing in Brain API proceed through explicit stages: candidate generation, policy filtering, contextual constraint application, ranking and selection, and plan synthesis. By separating hard constraints from preferences and by recording both inputs and outcomes, the control plane provides governed, adaptive, and inspectable routing semantics suitable for agentic and intent-driven systems.

\section{Execution Model}
\label{sec:8}

This section describes the execution model of Brain API and its interaction with heterogeneous backends. We define how execution plans are realized, how progress and outcomes are reported, and how failures and policy-driven adaptations are handled. The model emphasizes a clear boundary between decision-making and execution while preserving end-to-end observability and governance.

\subsection{Plan Realization}
\label{sec:8.1}

An execution begins when the control plane hands an execution plan to an appropriate execution adapter. The adapter binds abstract steps in the plan to concrete backend operations and translates control-plane constructs into backend-specific representations while preserving the structure and constraints of the plan.

In practice, backends span a wide range of execution substrates. A \textbf{Kubernetes}-backed adapter submits steps as Jobs or Pods with resource limits and node selectors derived from policy constraints; data residency requirements become node affinity rules, and cost ceilings become resource quotas \citep{burns2016kubernetes}. An \textbf{Argo Workflows} adapter maps a multi-step execution plan directly to a DAG or steps workflow, where each Brain API step becomes a workflow template invocation; branching and conditional steps in the plan translate to Argo's \texttt{when} expressions, and the resulting workflow manifest is submitted via the Argo API \citep{argoworkflows}. A \textbf{Temporal} adapter wraps each plan step as an Activity within a Workflow, leveraging Temporal's durable execution guarantees for long-running or failure-prone steps; policy-mandated retries and timeouts are expressed as Temporal retry policies and activity options \citep{temporal}. An \textbf{API gateway} adapter translates single-step plans into HTTP or gRPC calls with appropriate authentication and rate-limit headers derived from the decision artifact.

Execution plans may be linear, branching, or iterative, depending on the decomposition produced during decision-making. The execution model does not require a single global runtime; instead, it treats each backend as an autonomous executor that reports status and results back to the control plane through a uniform interface.

\subsection{Synchronous and Asynchronous Steps}
\label{sec:8.2}

The model supports both synchronous and asynchronous steps. Synchronous steps block the plan’s progression until completion and return a result or error. Asynchronous steps return a handle that can be polled or subscribed to for updates, allowing the plan to proceed concurrently or to wait on multiple outstanding operations. The execution adapter mediates these interactions and normalizes backend-specific notions of progress and completion into a common status model.

\subsection{State, Progress, and Checkpointing}
\label{sec:8.3}

During execution, the control plane maintains a logical execution state that records the status of each step, intermediate results, and any policy-relevant annotations (e.g., approvals obtained, constraints satisfied). This state enables checkpointing and partial recovery: if execution is interrupted, the control plane can resume from the last consistent checkpoint or trigger re-evaluation according to policy.

\subsection{Failure Semantics}
\label{sec:8.4}

Failures may occur at multiple levels, including step-level errors, backend unavailability, or violations of obligations attached during plan synthesis. The execution model distinguishes between \textbf{recoverable failures}, which permit retries or alternative choices, and \textbf{terminal failures}, which require aborting the plan and reporting an error to the caller.

Upon a recoverable failure, the control plane records the outcome and consults policy to determine the next action. Options include retrying the same step, selecting an alternative capability, or re-running the decision process with updated context. Terminal failures produce a final execution state and a corresponding decision outcome that is exposed through the observability plane.

\subsection{Policy-Driven Adaptation}
\label{sec:8.5}

Policies may mandate adaptive behavior at runtime, such as escalating to a higher-trust backend upon repeated failures, switching regions under latency constraints, or inserting approval steps before proceeding. The execution model supports such adaptations by allowing the control plane to pause execution, re-evaluate the decision with updated context, and either continue with a revised plan or terminate according to policy.

Crucially, adaptations are not implicit side effects: each adaptation results in a new or amended decision artifact linked to the prior one, preserving an auditable chain of control decisions.

\subsection{Concurrency and Coordination}
\label{sec:8.6}

Execution plans may express parallelism across independent steps. The execution model allows multiple steps to be dispatched concurrently, with coordination constructs (e.g., joins, barriers) expressed in the plan. The control plane tracks dependencies and ensures that downstream steps are activated only when their prerequisites have completed successfully or when policy allows alternative paths.

\subsection{Outcomes and Result Propagation}
\label{sec:8.7}

Upon completion, the execution produces an outcome that includes success or failure status, outputs of completed steps, and any policy-relevant annotations. This outcome is associated with the originating intent and decision artifact and is made available to clients through the execution interface. The outcome may also be persisted and fed back into context and memory providers to inform future decisions.

\subsection{Adapter Responsibilities and Backend Contracts}
\label{sec:8.8}

Regardless of the underlying backend, every execution adapter must satisfy the same contract with the control plane: (1) accept a typed execution plan and return a handle; (2) report step-level status updates via a uniform progress interface; (3) surface failures with enough context for the control plane to classify them as recoverable or terminal; and (4) return a typed outcome when execution completes. This contract is intentionally thin so that adapters for new backends can be added without modifying the core control plane.

For the running example---\emph{"Prepare a regulatory report for dataset D"}---the selected adapter would likely target an Argo Workflows or Temporal backend, depending on whether the plan is primarily a DAG of data-processing steps (favoring Argo's workflow-as-code model) or a long-running activity requiring durable retry semantics (favoring Temporal's event-driven workflow engine). A Kubernetes Job adapter could alternatively be used if report generation is a self-contained batch workload. In all three cases, Brain API has already enforced data residency and cost policies \emph{before} the adapter is invoked; the adapter simply realizes the compliant plan that the decision engine produced, without needing to re-implement governance logic.

\subsection{Summary}
\label{sec:8.9}

The execution model of Brain API provides a uniform, policy-aware abstraction over heterogeneous backends including Kubernetes, Argo Workflows, Temporal, and API-based services. By treating execution as the realization of explicit plans, recording progress and outcomes, and enabling controlled adaptation under policy, the model preserves a strict separation between decision-making and execution while supporting the needs of adaptive, agentic systems. Each backend is encapsulated behind a thin adapter contract, enabling incremental adoption and coexistence with established infrastructure.

\section{Observability and Governance}
\label{sec:9}

This section describes how Brain API provides observability and governance at the level of intents, decisions, and policies, rather than only at the level of resources or requests. We argue that decision-level visibility is essential for operating, auditing, and governing agentic systems, and we outline the mechanisms by which the control plane records, exposes, and enforces such visibility.

\subsection{Decision-Level Telemetry}
\label{sec:9.1}

Traditional observability stacks focus on metrics, logs, and traces associated with resource usage and request flows \citep{sigelman2010dapper}. Brain API extends this model by introducing \textbf{decision-level telemetry}. For each intent, the control plane records the inputs to the decision process (intent, policy versions, relevant context), the produced decision artifact, and the derived execution plan. This information is emitted as structured events that can be correlated with execution traces from underlying backends.

Decision-level telemetry enables operators to answer questions such as: which policies were applied to a given intent, which alternatives were considered, and why a particular plan was selected. This shifts observability from a purely operational view to a control-centric view that exposes the system’s reasoning process.

\subsection{Tracing Across Control and Data Planes}
\label{sec:9.2}

Brain API integrates control-plane and data-plane tracing by propagating identifiers from intents and decisions into execution steps. Each execution adapter attaches the corresponding intent and decision identifiers to backend-specific traces, enabling end-to-end correlation. As a result, operators can reconstruct not only \emph{what} actions were executed and \emph{where}, but also \emph{which decision} led to those actions and under \emph{which policies}.

This unified tracing model supports root-cause analysis that spans decision-making and execution, which is particularly important in adaptive systems where behavior may change in response to context or policy updates.

\subsection{Auditing and Compliance}
\label{sec:9.3}

Governance requirements in regulated or safety-critical environments often demand a durable audit trail of control decisions \citep{buneman2001why,cheney2009provenance}. Brain API treats decision artifacts as auditable records. Each artifact includes references to the intent, the policy set and versions in force, the context inputs used, and the selected execution plan. These records can be stored in an append-only log or ledger and queried for compliance, forensic analysis, or reporting.

Policies may also specify audit obligations, such as mandatory approval steps, segregation-of-duties constraints, or retention requirements for decision records. The control plane enforces these obligations during plan synthesis and execution, ensuring that governance is not merely observational but also prescriptive.

\subsection{Metrics at the Control Layer}
\label{sec:9.4}

In addition to traditional infrastructure metrics, Brain API exposes metrics at the control layer. Examples include intent throughput, decision latency, policy evaluation cost, frequency of re-evaluation, distribution of selected capabilities, and rates of fallback or re-routing. These metrics provide visibility into the behavior and performance of the decision process itself, enabling capacity planning and policy tuning.

\subsection{Policy Enforcement and Escalation}
\label{sec:9.5}

Governance is not limited to post hoc observation. Policies may require runtime enforcement actions, such as blocking certain classes of intents, requiring human approval before proceeding, or escalating execution to higher-trust environments. The control plane supports such actions by integrating policy checks into both the decision and execution phases. Violations or near-violations can trigger alerts, pauses, or re-evaluations according to policy.

\subsection{Explainability and Inspection}
\label{sec:9.6}

Because decisions are first-class artifacts, Brain API can expose explainability interfaces that summarize the factors contributing to a decision, including applicable policies, relevant context, and comparative scores of alternatives. While the exact form of explanation may vary by implementation, the architectural requirement is that sufficient information is preserved to support meaningful inspection by operators and auditors.

\subsection{Summary}
\label{sec:9.7}

By elevating observability and governance to the level of intents, decisions, and policies, Brain API provides a control-centric view of system behavior. This approach complements traditional infrastructure observability and enables auditing, compliance, and operational understanding of adaptive, agentic systems at scale.

\section{Use Cases}
\label{sec:10}

In this section, we illustrate how Brain API can be applied to representative scenarios that highlight the benefits of intent-first control, policy-governed decision-making, and decision-level observability. The examples are intentionally abstracted from specific products or implementations to emphasize the generality of the approach.

\subsection{Agentic Datasets as Policy-Governed Capabilities}
\label{sec:10.1}

\textbf{Traditional vs. agentic datasets.} In traditional systems, a dataset is a passive artifact: it stores data and responds to queries, but takes no part in deciding whether, how, or where it is used. In an agentic system, a dataset becomes an \emph{active capability node}: it exposes what operations are possible, what constraints govern its use, and what context signals are available to the decision engine---allowing the control plane to reason about the dataset alongside services, tools, and other agents.

\textbf{Definition.} An \emph{agentic dataset} is a dataset that registers with the Brain API control plane by exposing four elements: (1) \emph{capabilities} (operations that callers may request, e.g., query, transform, summarize, validate, export); (2) \emph{context signals} (dynamic observables such as freshness, size, sensitivity label, and workload); (3) \emph{policies} (governance constraints that restrict when and how each capability may be invoked); and (4) a \emph{decision interface} through which the control plane discovers, evaluates, and selects the dataset as a participant in an execution plan. This is precisely the structure of a \texttt{capability\_spec} (\S\ref{sec:6.8}), extended with a \texttt{kind: agentic\_dataset} marker and a structured \texttt{affordances} field.

\textbf{Dataset descriptor.} The following descriptor shows how the EU Regulatory Corpus would register with Brain API:

\begin{verbatim}
{
  "capability_id": "ds-regulatory-corpus-eu",
  "name": "EU Regulatory Corpus",
  "kind": "agentic_dataset",
  "affordances": ["query", "summarize", "validate", "export"],
  "input_types": ["intent", "filter_spec"],
  "output_types": ["tabular", "pdf", "json"],
  "trust_level": "elevated",
  "policies": ["eu_data_residency", "pii_restricted", "cost_ceiling_5usd"],
  "context_signals": {
    "freshness_hours": 4,
    "size_gb": 120,
    "sensitivity": "high"
  },
  "cost_model": { "type": "per-query", "usd": 0.50 },
  "data_residency": ["eu-west-1"]
}
\end{verbatim}

The \texttt{affordances} field enumerates \emph{dataset affordances}: the actions the dataset enables that are semantically meaningful to the decision engine. Unlike raw API endpoints, affordances are declared at the intent level---\texttt{summarize} means "produce a summary suitable for an intent," not a specific API call. This allows the decision engine to match a high-level intent against affordances across heterogeneous dataset types without requiring schema-level knowledge of each backend.

\textbf{Decision flow.} Consider the running example: \emph{"Prepare a regulatory report for dataset D."} With the descriptor above registered, the decision engine's evaluation proceeds as follows. First, candidate generation identifies \texttt{ds-regulatory-corpus-eu} (along with any other registered summarization capabilities) as relevant to the \texttt{summarize} affordance implied by the intent. Second, policy filtering removes non-compliant candidates: a US-hosted summarization service is eliminated by \texttt{eu\_data\_residency}; a cheaper cloud model is eliminated by \texttt{pii\_restricted}. Third, ranking scores the remaining candidates---including the dataset's own \texttt{summarize} affordance (reusing a prior result if freshness allows) and an on-premise summarization job that reads from the corpus---by cost, latency, and trust level. Fourth, the decision artifact records which candidate was selected and why, and the execution plan is dispatched. In this scenario, the dataset is not merely the input to a computation: it is one of the capability nodes in the decision graph, evaluated on equal footing with services and tools.

\textbf{Architecture relationship.} Brain API acts as the decision control plane; agentic datasets act as decision-aware data nodes within it. At the architectural level:

\begin{itemize}
\item Brain API provides intent parsing, policy enforcement, capability selection, and decision artifact production.
\item Agentic datasets provide structured capability declarations, context signals, and governance metadata that the control plane reasons over.
\item Existing data infrastructure (storage systems, query engines, data catalogs) is unchanged; the agentic dataset layer is a thin registration and context-exposure wrapper that makes existing datasets visible to the control plane.
\end{itemize}

This relationship is composable: a single execution plan may combine dataset nodes, model nodes, tool nodes, and workflow nodes, all governed uniformly by the same decision layer. Datasets become \emph{agentic} not by embedding control logic in storage systems, but by participating in a decision-centric control plane that governs how and when their capabilities are used.

\subsection{Agentic Data Processing Pipeline}
\label{sec:10.2}

Consider an organization that operates a heterogeneous data platform comprising batch processing jobs, interactive analytics services, and external data providers. A user submits the intent: \emph{“Analyze dataset D and produce a quality report.”} In a traditional system, this would require either a predefined workflow or bespoke orchestration logic embedded in application code.

With Brain API, the intent is submitted to the control plane, which evaluates policies related to data governance, cost ceilings, and execution locality. The capability registry advertises multiple processing options, including an on-premise cluster, a managed cloud service, and a specialized compliance-certified environment. Based on current context (e.g., cluster load, data sensitivity labels) and policies (e.g., data residency requirements), the decision engine selects an appropriate composition of capabilities and produces an execution plan. The observability plane records not only the execution outcomes, but also why a particular environment and toolchain were chosen.

\subsection{Policy-Constrained Tool-Augmented Agent}
\label{sec:10.3}

Consider an agent that assists analysts by invoking tools such as document search, data extraction, and report generation. The high-level intent might be \emph{"Prepare a regulatory summary for case X."} Policies specify that certain data sources require elevated trust levels and that external APIs may only be used if cost and privacy constraints are satisfied.

In this scenario, Brain API mediates tool selection. Candidate tools are generated based on semantic compatibility with the intent. Policies filter out tools that do not meet trust or compliance requirements, and ranking prefers tools with lower estimated cost and latency. The resulting decision and plan are recorded, enabling auditors to verify that only approved tools and data sources were used. If a tool invocation fails or violates a constraint, policy may trigger re-evaluation and selection of an alternative tool.

\subsection{Multi-Backend Routing and Fallback}
\label{sec:10.4}

Consider a service that can be fulfilled by multiple backends with different cost and performance profiles, such as a local deployment, a regional cloud service, and a globally replicated service. The intent \emph{“Process request R under latency L and cost C constraints”} is submitted. Policies encode budget limits and availability requirements, while context provides real-time load and health information.

Brain API generates candidates for each backend, filters them by policy, and ranks them based on current conditions. The selected plan may initially target the local deployment. If execution fails or latency thresholds are exceeded, policy may require a fallback to a regional or global backend. Each such transition is driven by explicit re-evaluation and produces a new decision artifact, preserving an auditable trail of routing choices.

\subsection{Semantic Memory and Result Reuse (Preview)}
\label{sec:10.5}

In some domains, repeated or semantically similar intents may benefit from reusing prior results. For example, an intent such as \emph{“Summarize dataset D with parameters P”} may be close to a previously executed intent. A memory or caching subsystem can expose signals indicating similarity and freshness of past outcomes.

Within Brain API, such signals appear as context inputs to the decision process. Policies may allow reuse only if similarity exceeds a threshold and results are sufficiently fresh. The decision engine can then prefer a plan that retrieves a prior result over recomputing it, or it can select recomputation if constraints are not met. Importantly, this behavior is expressed and governed at the control-plane level, and the decision artifact records whether reuse or recomputation was chosen and why. A dedicated semantic memory plane can further specialize this pattern, which we leave to future work.

\subsection{Summary}
\label{sec:10.6}

These use cases demonstrate how Brain API generalizes control over heterogeneous, adaptive, and policy-sensitive systems. By making intent, policy, and decision-making explicit, the control plane supports scenarios that would otherwise require ad hoc orchestration logic, while preserving observability and governance across diverse execution environments.

\section{Evaluation}
\label{sec:11}

\begin{figure}[!t]
  \centering
  \begin{tikzpicture}[x=1cm, y=1cm,
    tile/.style={minimum width=3.85cm, minimum height=#1, inner sep=0pt},
    tt/.style={anchor=north west, font=\scriptsize\bfseries, inner sep=1pt},
    tr/.style={anchor=north east, font=\scriptsize, text=black!55, inner sep=1pt},
    big/.style={anchor=north west, font=\Large\bfseries, inner sep=1pt},
    body/.style={anchor=north west, font=\scriptsize, text width=3.55cm, align=left, inner sep=1pt}]

  \node[ev, minimum width=16cm, minimum height=1.95cm] at (8,6.25) {};
  \node[tt] at (0.15,7.12) {Expressiveness of the rule grammar};
  \node[tr] at (15.85,7.12) {\S\ref*{sec:11.2}};
  \fill[black!80] (0.15,6.3) rectangle (1.12,6.7);
  \fill[blue!55!black] (1.12,6.3) rectangle (9.28,6.7);
  \fill[black!10] (9.28,6.3) rectangle (15.85,6.7);
  \draw[black!40, line width=0.4pt] (0.15,6.3) rectangle (15.85,6.7);
  \node[anchor=north west, font=\scriptsize, inner sep=1pt, text width=15.6cm, align=left] at (0.15,6.2)
    {\textbf{3 of 49} policies as first specified \qquad
     \textcolor{blue!55!black}{\textbf{28 of 49}} with bounded quantification \qquad
     \textbf{21} need aggregation or general regular expressions\\[1pt] {\color{black!55}reported, rather than the boundary adjusted to improve the figure}};

  \foreach \x/\t/\r in {2.025/{Policy-stage conformance}/{\S\ref*{sec:11.5}},
                        6.075/{Auditability}/{\S\ref*{sec:11.3}},
                        10.125/{Selection}/{\S\ref*{sec:11.4}},
                        14.175/{Second domain: Cedar}/{\S\ref*{sec:11.6}}} {
    \node[ev, tile=2.75cm] at (\x,3.55) {};
    \node[tt] at ($(\x,4.83)+(-1.8,0)$) {\t};
    \node[tr] at ($(\x,4.83)+(1.8,0)$) {\r};
  }
  \node[big] at (0.25,4.3) {42 {\normalsize\mdseries of 42}};
  \node[body] at (0.25,3.62) {agree with the library's published verdicts: 19 admit, 23 deny. Admitting everything would score 45 per cent.};
  \node[anchor=west, font=\scriptsize] at (4.3,4.05) {artifact};
  \node[anchor=west, font=\scriptsize] at (4.3,3.6) {admission log};
  \foreach \q/\a/\l in {1/1/1, 2/1/0, 3/1/0, 4/1/1, 5/1/0, 6/1/0} {
    \pgfmathsetmacro{\px}{5.85 + 0.24*\q}
    \ifnum\a=1 \fill[green!45!black] (\px,4.05) circle (0.085); \else \draw[black!45] (\px,4.05) circle (0.085); \fi
    \ifnum\l=1 \fill[green!45!black] (\px,3.6) circle (0.085); \else \draw[black!45] (\px,3.6) circle (0.085); \fi
  }
  \node[anchor=west, font=\scriptsize\bfseries] at (7.45,4.05) {6/6};
  \node[anchor=west, font=\scriptsize\bfseries] at (7.45,3.6) {2/6};
  \node[body] at (4.3,3.25) {questions Q1--Q6, answerable from the record alone; replay re-derived 10 of 10 decisions};
  \node[big] at (8.35,4.3) {20 {\normalsize\mdseries of 20}};
  \node[body] at (8.35,3.62) {trials selected the alternative the corpus labels compliant, in one pass: ten pairs, both orderings.};
  \node[big] at (12.4,4.3) {78 {\normalsize\mdseries of 81}};
  \node[body] at (12.4,3.62) {agree with Cedar's reference implementation: 71 of 71 denials, 7 of 10 allows. None admitted that Cedar denies.};

  \node[anchor=west, font=\scriptsize\bfseries] at (0,1.95) {Not evaluated};
  \foreach \x/\t/\b in {1.35/{Candidate generation}/{runs, not measured},
                        4.05/{Context-signal stage}/{design claim},
                        6.75/{Ranking stage}/{design claim},
                        9.45/{Plan synthesis}/{runs, not measured},
                        12.15/{Latency under load}/{unquantified},
                        14.85/{Execution}/{outside the prototype}} {
    \node[de, tile=0.95cm, minimum width=2.55cm] at (\x,1.2) {};
    \node[anchor=north west, font=\scriptsize\bfseries, text=black!65, inner sep=1pt, text width=2.4cm, align=left] at ($(\x,1.6)+(-1.2,0)$) {\t};
    \node[anchor=north west, font=\scriptsize, text=black!55, inner sep=1pt, text width=2.4cm, align=left] at ($(\x,1.23)+(-1.2,0)$) {\b};
  }

  \figlegend{0,0.3}
\end{tikzpicture}
  \caption{What \S\ref{sec:11} measured, and what it did not. Every figure is one the section states. The dashed row repeats the abstract: candidate generation, context signals, ranking and plan synthesis are not measured, nor is decision latency under load; execution lies outside a prototype of the decision layer.}
  \label{fig:evaluation}
\end{figure}
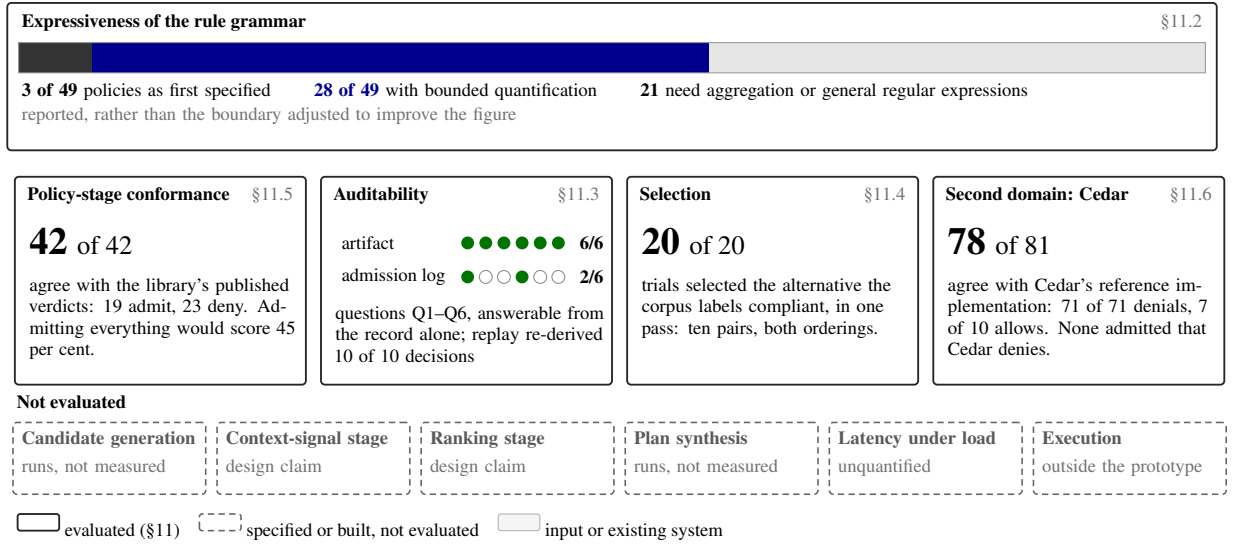

We evaluate a prototype of the decision layer against external policy corpora. The prototype implements the five-stage pipeline of \S\ref{sec:7} in approximately 1,600 lines of Rust and is exercised by the experiments below. Figure~\ref{fig:evaluation} summarizes the results, and what was not evaluated.

\subsection{Method and corpus}
\label{sec:11.1}

The evaluation constraint that governs this section is that the workload must not define the property it measures. We therefore take both the policies and the expected outcomes from a third party: the OPA Gatekeeper constraint library \citep{gatekeeperlibrary}, a set of Kubernetes admission policies in production use. It comprises 49 constraint templates, 69 parameterized constraint instances, and 180 sample objects, each of which the library's own test suite labels with the verdict it expects:

\begin{verbatim}
cases:
- name: example-allowed
  object: samples/ingress-https-only/example_allowed.yaml
  assertions:
  - violations: no
- name: example-disallowed
  object: samples/ingress-https-only/example_disallowed.yaml
  assertions:
  - violations: yes
\end{verbatim}

We author the encoding of each policy into the rule grammar of \S\ref{sec:6.3}. We author neither the sample objects nor the verdicts. An incorrect encoding therefore surfaces as disagreement with the corpus, not as a favorable measurement --- and, as \S\ref{sec:11.2} records, this is precisely how one deficiency in the rule grammar was found.

The domain is a deliberate fit rather than a convenient one: Kubernetes admission control \emph{is} pre-execution policy gating, which is the regime this paper argues about. A second and deliberately dissimilar domain --- Cedar authorization --- is evaluated in \S\ref{sec:11.6}.

We considered and rejected an alternative corpus. The Cedar policy language \citep{cedar2024} ships 7,600 integration tests, each carrying an authorization decision and the identity of the policies that produced it --- an appealing oracle, since the latter is ground truth for artifact content. On inspection the corpus is generated by coverage-guided fuzzing, and its policies are correspondingly degenerate (\texttt{permit(principal, action, resource) when \{ true \}}) or adversarial. It is well suited to testing an implementation of Cedar and unsuited to standing in for governance policy, so we did not use it.

\subsection{Expressiveness: what real policies require}
\label{sec:11.2}

Our first result concerns the rule grammar itself, and it is negative.

As originally specified, a policy rule is a comparison of one field against one value, optionally guarded by a condition. Measured against the corpus, that grammar expresses \textbf{3 of 49 policies}. The reason is uniform: real governance policies quantify over collections. The commonest shape in the corpus is

\begin{verbatim}
container := input.review.object.spec.containers[_]
not strings.any_prefix_match(container.image, input.parameters.repos)
\end{verbatim}

--- "every container's image must come from an allowed repository" --- which no scalar comparison can express. Across the corpus, 45 of 49 templates quantify and 19 aggregate.

We therefore extended the grammar with \textbf{bounded quantification}: a rule may range over a collection path with an \texttt{All} or \texttt{Any} quantifier, comparing a field of each element. With list-valued membership and a negated match operator, coverage rises to \textbf{28 of 49}. Aggregation (counting, summing, arithmetic over resource quantities) and general regular expressions remain outside the grammar; we report the 21 policies that need them rather than adjusting the boundary to improve the figure.

A second deficiency was found by the corpus rather than by inspection. An initial conformance run agreed with 37 of 42 cases, and all five disagreements were exemption cases. The library's shared \texttt{exempt\_container} module excuses a container whose image appears on an exemption list, so the policy is properly read as "deny if any container is privileged \emph{and} its image is not exempt" --- a conjunction over the \emph{same} quantified element. The rule-level condition of \S\ref{sec:6.3} cannot express this, because it is evaluated against the capability rather than against the element under examination. We added per-element conditions accordingly.

We draw attention to this because the sequence matters more than the fix. Both extensions were forced by an external corpus; neither was anticipated when the model was specified. A design validated only against scenarios of its authors' construction would have retained both defects, and the second was invisible to inspection.

\subsection{Decision artifacts and auditability}
\label{sec:11.3}

We now test the first of the two empirical claims of \S\ref{sec:1}: that existing systems do not produce a decision artifact.

The comparison is against admission control as it is actually deployed --- a checker that receives one proposed object, admits or rejects it, and emits a log record. We grant the baseline the violation messages it genuinely produces, since Gatekeeper does explain its rejections; understating it would make the comparison worthless.

We pose six questions that an auditor asks after the fact, and ask which can be answered \textbf{from a record alone}, without re-running the engine and without access to the policy set:

\begin{table}[h]
\centering
\small
\begin{tabular}{|l|l|l|}
\hline
Question & Decision artifact & Admission log \\
\hline
Q1 Was the request admitted? & yes & yes \\
Q2 Which capability served it? & yes & no \\
Q3 What else was considered? & yes & no \\
Q4 Why was each rejected alternative rejected? & yes & yes \\
Q5 Was a compliant alternative available? & yes & no \\
Q6 Can the decision be re-derived and checked? & yes & no \\
\hline
\end{tabular}
\end{table}

Averaged over the corpus decisions, the decision artifact answers 6.0 of 6 and the admission log 2.0 of 6. The four questions the log cannot answer follow from what it holds: one proposal and one verdict. Two concern alternatives, which an admission controller is never offered and so cannot report; the other two ask which capability was chosen from among several, and whether the decision can be reproduced from the record, neither of which a single verdict carries.

Q6 is the strongest of the six and the only one whose evaluation is immune to any choice of ours, because it judges nothing --- it merely reproduces. Given the artifact alone, we re-derive the selection (the top-ranked candidate that passed policy) and check it against the decision actually taken. This succeeded for \textbf{10 of 10} decisions. An auditor can therefore verify a Brain API decision without the engine, the policies, or the original inputs.

Figure~\ref{fig:anatomy} shows one of these records against the seven contents that \S\ref{sec:4.1} defines.

\begin{figure}[!ht]
  \centering
  \begin{tikzpicture}[x=1cm, y=1cm,
    row/.style={minimum width=16cm, minimum height=#1, inner sep=0pt},
    row/.default=0.62cm,
    lab/.style={anchor=west, font=\scriptsize\bfseries, inner sep=1pt},
    st/.style={anchor=west, font=\scriptsize, inner sep=1pt},
    val/.style={anchor=west, font=\scriptsize, text width=9.7cm, align=left, inner sep=1pt}]
  \def\recorded{\textcolor{green!45!black}{recorded}}
  \def\inpart{\textcolor{blue!55!black}{in part}}
  \def\notrec{\textcolor{black!50}{not recorded}}

  \draw[line width=1.1pt, black!85, rounded corners=3pt] (0,0.62) rectangle (16,8.75);
  \fill[black!5, rounded corners=3pt] (0.03,8.1) rectangle (15.97,8.72);
  \draw[black!30, line width=0.4pt] (0,8.1) -- (16,8.1);
  \node[anchor=west, font=\scriptsize\bfseries] at (0.25,8.42) {DECISION ARTIFACT};
  \node[anchor=west, font=\footnotesize\ttfamily] at (3.25,8.42) {intent deploy-001};
  \node[anchor=east, font=\scriptsize\ttfamily, text=black!60] at (15.8,8.42) {schema brainapi.decision-artifact/v1};

  \foreach \y in {7.45,6.8,6.1,5.4,4.35,3.3} { \draw[black!18, line width=0.4pt] (0.15,\y) -- (15.85,\y); }

  \node[lab] at (0.25,7.77) {1\enspace Intent reference};
  \node[st]  at (4.1,7.77) {\inpart};
  \node[val] at (6.0,7.77) {\texttt{deploy-001}: the identifier; the intent's parameters and preferences are not recorded};

  \node[lab] at (0.25,7.12) {2\enspace Policy versions};
  \node[st]  at (4.1,7.12) {\inpart};
  \node[val] at (6.0,7.12) {\texttt{allow-privilege-escalation v1.0}, carried inside the violation text rather than as a field of its own};

  \node[de, minimum width=15.6cm, minimum height=0.5cm] at (8,6.45) {};
  \node[lab, text=black!55] at (0.25,6.45) {3\enspace Context snapshot};
  \node[st]  at (4.1,6.45) {\notrec};
  \node[val, text=black!55] at (6.0,6.45) {the prototype records no context signals};

  \node[lab] at (0.25,5.75) {4\enspace Candidate set};
  \node[st]  at (4.1,5.75) {\recorded};
  \node[val] at (6.0,5.75) {\texttt{deploy.variant-a}, \texttt{deploy.variant-b}, each with its policy result};

  \node[lab] at (0.25,4.88) {5\enspace Selected plan};
  \node[st]  at (4.1,4.88) {\inpart};
  \node[val] at (6.0,4.88) {\texttt{deploy.variant-b}: the capability, with its parameters (the Pod object); the plan's steps are not recorded};

  \node[lab] at (0.25,3.83) {6\enspace Rejected alternatives};
  \node[st]  at (4.1,3.83) {\recorded};
  \node[val] at (6.0,3.83) {\xmark\ \texttt{deploy.variant-a}: \texttt{policy \textquotesingle{}allow-privilege-escalation\textquotesingle{} v1.0: denied by rule on \textquotesingle{}any of \textquotesingle{}capability.\allowbreak spec.\allowbreak ephemeralContainers\textquotesingle{}.\allowbreak securityContext.\allowbreak allowPrivilegeEscalation\textquotesingle{}}};

  \node[lab] at (0.25,2.9) {7\enspace Decision rationale};
  \node[st]  at (4.1,2.9) {\recorded};
  \node[val] at (6.0,2.9) {\texttt{Selected \textquotesingle{}deploy.variant-b\textquotesingle{} (agent 2, score: 2.00)}};

  \draw[black!30, line width=0.4pt] (0,2.45) -- (16,2.45);
  \node[anchor=west, font=\scriptsize\bfseries] at (0.25,2.12) {What this record answers (\S\ref*{sec:11.3})};
  \node[anchor=west, font=\scriptsize, text width=15.4cm, align=left] at (0.25,1.35)
    {\textbf{Audit}\enspace Q1 was it admitted? \enspace Q2 which capability served it? \qquad
     \textbf{Compare}\enspace Q3 what else was considered? \enspace Q5 was a compliant alternative available?\\[2pt]
     \textbf{Explain}\enspace Q4 why was each rejected alternative rejected? \qquad
     \textbf{Replay}\enspace Q6 can the decision be re-derived and checked?\\[3pt]
     \textcolor{green!45!black}{\textbf{6 of 6}} from this record \qquad
     \textbf{2 of 6} (Q1, Q4) from the admission log for the same case, which saw only \texttt{deploy.variant-a}};

  \figlegend{0,0.15}
\end{tikzpicture}
  \caption{Anatomy of a decision artifact: one of the ten decisions of \S\ref{sec:11.3}, as the prototype recorded it, for the Gatekeeper policy \texttt{allow-privilege-escalation}. Rows are the seven contents \S\ref{sec:4.1} defines. The record carries the candidate set, the rejected alternative with its cause, and the rationale; it carries the intent by identifier only, policy versions only inside violation text, and the selected capability without its plan steps; it carries no context snapshot. The six audit questions of \S\ref{sec:11.3} are answerable from what it does record.}
  \label{fig:anatomy}
\end{figure}

\subsection{Selection among alternatives}
\label{sec:11.4}

The second claim requires more care than it is usually given. Admission control also runs before execution, so pre-execution timing is \emph{not} what distinguishes intent-level enforcement. The distinction is that an admission controller adjudicates one proposed object, whereas the decision layer selects among alternatives.

The corpus supplies the alternatives without our inventing them. Ten policies ship both a compliant and a non-compliant sample object for the same constraint --- third-party-authored variants of the same capability, with third-party labels. We present both as candidate capabilities for a single intent, under both orderings so that nothing depends on which is offered first, and give the candidates neutral positional names so that no label is visible to the engine.

Brain API selected the capability the corpus labels compliant in \textbf{20 of 20 trials}, in one pass, with no failures to decide.

The baseline is handed one proposal. Where that proposal is compliant it is admitted; where it is not, it is rejected with an explanation, and reaching a compliant execution requires a further admission round trip --- and only if something upstream happens to offer the other variant, since the log names no alternative. Over the same trials this averages 1.50 round trips per intent against 1.00, with 10 rejections from which the record affords no route forward.

We state plainly what this does and does not show. It does not show that Brain API enforces policy more accurately than Gatekeeper; Gatekeeper is correct here by construction, and evaluates a strictly larger policy language than ours. What it shows is that adjudicating a proposal and selecting among alternatives are different operations, and that a system built for the former cannot perform the latter however well it performs its own.

\subsection{Conformance of the policy stage}
\label{sec:11.5}

Underlying both experiments is the question of whether the policy stage decides correctly at all. Over the encodable fragment, the prototype agrees with the corpus's published verdicts on \textbf{42 of 42 cases}, comprising 19 admit and 23 deny. We report the class balance because without it the figure would be uninterpretable: an encoding that admitted everything would score 45 per cent.

Nine cases were not evaluated, for two identified reasons, which we report rather than omit. One policy requires case-insensitive comparison; another requires quantification nested two levels deep, over volume mounts within containers, where the grammar quantifies over one.

Conformance establishes that the policy stage is \emph{sound}, not that it is superior. It is a precondition for the results of \S\ref{sec:11.3} and \S\ref{sec:11.4} rather than a claim in its own right.

\subsection{A second domain: authorization}
\label{sec:11.6}

Everything above is Kubernetes admission control. To test whether any of it generalizes we repeated the exercise on a domain chosen to differ in vendor, in paradigm and in subject matter: the Cedar policy language, whose published example policies are business scenarios --- document sharing, streaming entitlements, tax-document access --- and whose model is principal/action/resource authorization rather than object admission.

Cedar's examples ship policies and entities but no expected decisions, so we obtained labels by differential testing against the Cedar CLI, the reference implementation of the language. Requests are the exhaustive principal $\times$ action $\times$ resource cross-product of each example's entity file; we choose which questions to ask, and a third-party implementation answers every one of them.

The domain repaid the effort before a single request was evaluated, by exposing three gaps that Gatekeeper could not --- because Gatekeeper happens to share the shape the model was built around.

\textbf{Polarity.} Cedar permits a request only when some \texttt{permit} matches and no \texttt{forbid} does. Our model was default-allow throughout: \texttt{PolicyAction::Allow} produced no violation and therefore changed no outcome, making it inert. Under default-allow semantics every Cedar policy would admit whatever its \texttt{forbid} statements failed to catch --- a silent and total inversion of intent. Admission control is default-allow and authorization is default-deny; a control plane claiming to unify governance across heterogeneous backends must represent both, and we had assumed one.

\textbf{Conjunctive guards.} A Cedar \texttt{permit} scope conjoins principal type, action and resource type before any \texttt{when} or \texttt{unless} clause. Our rule form carried a single optional guard, which cannot encode even the simplest such policy. Guards are now a conjunctive list.

\textbf{Cross-entity comparison}, which we did \emph{not} fix. Cedar routinely compares one entity's attributes against another's:

\begin{verbatim}
principal.assigned_orgs.contains(
  { organization: resource.owner.organization,
    serviceline: resource.serviceline, location: resource.location })
\end{verbatim}

Our rules compare a field against a constant and have no variable binding across entities. This is the dominant blocker --- 24 of 35 statements --- and closing it means introducing an expression language, not extending a rule form. We record it as the boundary of the design rather than crossing it.

By static triage, 8 of 35 policy conditions (23 per cent) fall inside the extended grammar, against 28 of 49 for Gatekeeper. On the encodable fragment of the streaming-entitlement example, over 81 requests:

\begin{table}[h]
\centering
\small
\begin{tabular}{|l|l|}
\hline
Measure & Result \\
\hline
Agreement with the Cedar CLI & 78 / 81 \\
--- on Cedar's denials & 71 / 71 \\
--- on Cedar's allows & 7 / 10 \\
Requests we admit that Cedar denies & \textbf{0} \\
Requests we deny that Cedar admits & 3 \\
\hline
\end{tabular}
\end{table}

We report agreement on allows separately because the aggregate flatters: the request set is dominated by denials, which a default-deny encoding reproduces for free. All three disagreements are requests permitted by time-gated statements we could not encode, and in every case the divergence is conservative --- the prototype never admits a request the reference implementation refuses.

The honest summary of this domain is mixed, and we prefer it to a second easy result. The decision layer's structure transfers: the same engine, extended twice, reproduces a different vendor's authorization semantics on the fragment it can express, and errs safe where it cannot. The rule \emph{language} transfers poorly: authorization policy is substantially about relations between entities, and a grammar of field-against-constant comparisons reaches under a quarter of it.

\subsection{Threats to validity}
\label{sec:11.7}

\textbf{Domain coverage.} Two domains are better than one, but both are policy adjudication. Neither is the multi-step regulated-reporting workflow used as this paper's running example, and the mapping to that scenario remains argued rather than demonstrated.

\textbf{Differential labels.} In the second domain the labels come from a reference implementation rather than from human-authored expectations. This is standard practice and it removes our judgement from the answers, but it inherits whatever that implementation does --- including on requests its own authors never considered.

\textbf{Encoding authorship.} We wrote the policy encodings. A systematically favorable encoding is conceivable in principle; conformance against third-party verdicts is what constrains it, and the 37-of-42 run that exposed the exemption defect is evidence that the constraint binds.

\textbf{Scale.} Ten alternative-pairs and 42 conformance cases are a small evaluation. The results are consistent and the mechanisms are structural rather than statistical, but we do not claim tight estimates from them.

\textbf{Stages not evaluated.} The context-signal and ranking stages are untested here, because pass/fail admission decisions supply no ranking ground truth. Claims about those stages remain design claims. Candidate generation and plan synthesis run in every experiment, but no experiment measures them.

\section{Discussion and Limitations}
\label{sec:12}

In this section, we discuss the implications of the Brain API design, analyze its limitations, and outline trade-offs inherent in introducing an intent- and policy-aware control plane. While the proposed approach provides a unifying abstraction for governed, adaptive systems, it also introduces new considerations in scalability, complexity, and operational practice.

\subsection{Scalability and Performance Overhead}
\label{sec:12.1}

Introducing a decision layer adds latency and computational cost relative to direct invocation of execution backends. Policy evaluation, candidate generation, and ranking incur overhead that must be managed carefully in high-throughput or latency-sensitive environments. Caching of decisions, incremental re-evaluation, and tiered policy checks can mitigate these costs, but they introduce additional engineering complexity. In practice, the suitability of Brain API for a given workload depends on the acceptable trade-off between control-plane sophistication and end-to-end latency.

\subsection{Policy Complexity and Maintainability}
\label{sec:12.2}

Elevating policy to a first-class control-plane primitive improves governance but also raises challenges in policy authoring, validation, and evolution. Large organizations may accumulate complex policy sets with interacting constraints and exceptions. Without careful tooling, such policies risk becoming difficult to reason about or debug. While Brain API provides the structural hooks for policy management, effective policy engineering practices and supporting tools remain essential and are outside the scope of this paper.

To ground this concern concretely, consider the following three policies that a regulated financial organization might define independently:

\begin{itemize}
\item \textbf{P1 (Data Residency):} \emph{"Data classified as \texttt{eu-personal} must not be processed outside \texttt{eu-west-1} or \texttt{eu-central-1}."}
\item \textbf{P2 (Trust Level):} \emph{"Processing of data classified as \texttt{sensitive-financial} must use a capability with \texttt{trust\_level = elevated}. The only registered elevated-trust capability is deployed in \texttt{us-east-1}."}
\item \textbf{P3 (Cost Gate):} \emph{"Any capability invocation with an estimated cost exceeding \$5.00 must be pre-approved by a human reviewer."}
\end{itemize}

A dataset that carries both \texttt{eu-personal} and \texttt{sensitive-financial} labels will produce an unsatisfiable intent under P1 and P2 taken together: P1 restricts execution to EU regions, while P2 requires a capability that exists only in \texttt{us-east-1}. The filtered candidate set $C' = \emptyset$, and the system signals an unsatisfiable intent. Without static analysis of the joint policy set, this conflict is only discovered at request time---potentially surfacing as an opaque failure in a production environment.

A second class of subtle interaction involves ordering and precedence. Consider adding:

\begin{itemize}
\item \textbf{P4 (Result Reuse):} \emph{"Prefer cached results when semantic similarity exceeds 0.85."}
\item \textbf{P5 (Regulatory Override):} \emph{"Intents with goal \texttt{produce-regulatory-summary} must always use fresh computation; cached results are not admissible."}
\end{itemize}

P4 and P5 interact: for regulatory intents, P5 overrides P4. If the policy engine applies P4 first and P5 is evaluated later as an additional filter, the outcome may depend on evaluation order---a fragile property that makes the system difficult to reason about and audit.

These examples illustrate two canonical failure modes: \textbf{mutual exclusion} (no candidate satisfies all policies simultaneously) and \textbf{implicit ordering dependence} (correct behavior depends on evaluation order among policies). Both are practically significant and require dedicated tooling---conflict detection, satisfiability checking, and policy simulation---to manage at scale. Brain API's architecture makes these failure modes observable (the decision artifact records $C' = \emptyset$ and which policies filtered which candidates), but detecting and resolving them proactively is a policy engineering problem that the control plane cannot solve alone.

\subsection{Debuggability and Explainability}
\label{sec:12.3}

Decision artifacts and decision-level telemetry improve transparency, but they also expose the need for clear explanation mechanisms. As decision functions incorporate multiple criteria, context signals, and historical data, the rationale for a particular choice may become non-trivial to summarize. Providing concise, operator-friendly explanations is an important practical concern. Our architecture preserves the information required for explanation, but the presentation and user experience of such explanations are left to implementation.

\subsection{Consistency and Distributed State}
\label{sec:12.4}

Brain API assumes access to policies, capability metadata, and context that may be distributed or replicated. Ensuring consistency across these inputs is a non-trivial problem in large-scale systems. Stale context or divergent policy views can lead to suboptimal or inconsistent decisions. Techniques from distributed systems, such as versioning, snapshotting, and eventual consistency with reconciliation, can be applied, but they introduce additional design choices and trade-offs.

\subsection{Scope Boundaries}
\label{sec:12.5}

Brain API is not intended to replace existing orchestration, workflow, or scheduling systems. Instead, it composes with them by providing a higher-level control plane. For workloads that are fully static, latency-critical, or trivially orchestrated, the additional abstraction may not be justified. Similarly, Brain API does not prescribe specific implementations for planners, memory systems, or policy languages, which means that the quality of decisions ultimately depends on the components integrated beneath the control plane.

\subsubsection{Security and Trust Considerations}
\label{sec:12.5.1}

Brain API introduces a control plane through which all intent-to-execution decisions flow, which makes it both a natural enforcement point and an attractive attack surface. We outline the primary threat categories and the architectural mitigations the design provides.

\textbf{Threat 1: Unauthorized or malicious intent submission.} A malicious entity could submit intents that attempt to invoke sensitive capabilities, exfiltrate data, or exhaust resources. The intent ingress is the first line of defense: it is responsible for authenticating the caller and validating the intent's schema and authorization scope before the intent reaches the decision engine. The capability registry augments this by attaching trust-level requirements to each capability (as illustrated by the \texttt{trust\_level} field in \S\ref{sec:6.8}). A caller with insufficient trust cannot be routed to elevated-trust capabilities; attempting to do so results in either policy filtering eliminating all compliant candidates (\S\ref{sec:7.3}) or an \emph{unsatisfied intent} signal rather than silent execution.  

\textbf{Threat 2: Policy bypass via capability manipulation.} An adversary with write access to the capability registry could register a malicious capability with falsely declared properties (e.g., an inflated trust level), causing the decision engine to select it for sensitive intents. The mitigation is to treat the capability registry as a trusted, access-controlled system: only authorized operators may register or update capabilities, and policy evaluation occurs over the declared metadata. Implementations should additionally consider out-of-band validation or attestation of capability metadata.

\textbf{Threat 3: Context poisoning.} Context providers supply inputs that influence ranking and, in some cases, filtering. A compromised context provider could manipulate signals (e.g., reporting false load or sensitivity labels) to cause the decision engine to select a suboptimal or non-compliant plan. The architecture mitigates this by (a) treating context signals as advisory inputs to ranking, not as overrides to hard policy constraints (\S\ref{sec:7.5.1}, Score independence property), and (b) recording all context references in the decision artifact, creating an auditable record that can detect anomalous context values post hoc.

\textbf{Threat 4: Decision artifact tampering.} If decision artifacts are mutable, an adversary could retroactively alter the recorded rationale to conceal a governance violation. The observability plane should persist decision artifacts as append-only records with cryptographic integrity guarantees, making tampering detectable.

\textbf{Scope note.} A full cryptographic security analysis, access-control model, and threat model formalization are outside the scope of this design proposal and represent important directions for future work. The intent here is to demonstrate that the architecture has natural enforcement points for the most prominent threat categories and to flag the areas where deployment-time security engineering is required.

\subsection{Summary}
\label{sec:12.6}

The Brain API approach introduces a principled control layer for intent-driven, policy-governed systems, but it also shifts complexity into the control plane. Understanding and managing the trade-offs in performance, policy engineering, distributed state, and security is essential for practical adoption. These limitations highlight both the challenges and the opportunities for future work in control-plane design for agentic systems.

\subsection{Evaluation Criteria}
\label{sec:12.7}

The evaluation in \S\ref{sec:11} covers policy filtering and selection; the questions below define what a fuller evaluation would address. Grounding these questions now serves two purposes: it makes the design commitments testable, and it provides a concrete experimental agenda for follow-on work.

\textbf{EQ1 --- Decision latency.} What is the end-to-end latency of the control-plane decision cycle (intent submission $\to$ decision artifact $\to$ execution dispatch), and how does it scale with the number of candidate capabilities, the complexity of the policy set, and the depth of the context graph? A practical threshold is that control-plane overhead should remain below 10\% of median backend execution latency for the workloads it governs. Micro-benchmarks should isolate candidate generation, policy filtering, and scoring as independent contributors.

\textbf{EQ2 --- Policy compliance rate.} What fraction of executed plans satisfy all applicable policies, as verified by post-hoc audit of the decision artifact against the policy set? Under correct operation the compliance rate should be 1.0; deviations indicate implementation bugs or policy ambiguity. A secondary question is the false-rejection rate: how often does the policy filter eliminate a candidate that a human auditor would have approved? These two metrics jointly characterize the precision and recall of the policy enforcement layer.

\textbf{EQ3 --- Routing optimality.} How close is the selected execution plan to the oracle-optimal plan with respect to the declared objective function (e.g., minimum cost, minimum latency, or maximum capability score)? Optimality gap can be measured against an exhaustive search baseline on small candidate sets, and approximation quality can be characterized as candidate-set size and policy complexity grow. This evaluation question assesses whether the scoring and ranking mechanism ($s: C' \times \Gamma \rightarrow \mathbb{R}$ and selection function $F$) produce decisions that are coherent with the declared objectives.

\textbf{EQ4 --- Cost efficiency.} In domains where execution has measurable cost (compute, API calls, data transfer), does Brain API's policy-governed selection produce lower total cost than baseline approaches such as round-robin dispatch, static routing rules, or unconstrained LLM tool selection? Cost efficiency should be measured under budget-constraint policies and compared against a policy-free baseline to quantify the value of intent-level governance.

\textbf{EQ5 --- Decision artifact utility.} Can the decision artifact support after-the-fact auditing, debugging, and policy refinement? Qualitative evaluation should assess whether practitioners can (a) determine from the artifact alone why a given candidate was selected or rejected, (b) replay the decision under a modified policy without re-running execution, and (c) detect policy conflicts from artifact analysis. This question evaluates the observability and governance value of the decision layer independently of its routing quality.

\textbf{EQ6 --- Policy evaluation overhead and capability selection complexity.} What is the computational cost of policy evaluation as a function of policy set size $|P|$, candidate set size $|C'|$, and the structural complexity of individual policies (e.g., conjunctive depth, number of context attributes referenced)? Similarly, how does candidate generation complexity grow with registry size $|C|$? In the worst case, candidate generation is $O(|C|)$ and policy filtering is $O(|C| \cdot |P|)$; empirical evaluation should characterize constant factors and identify the regime at which caching or incremental evaluation becomes necessary. These measurements inform practical deployment decisions about registry sharding, policy indexing, and tiered evaluation strategies.

\textbf{EQ7 --- Scaling characteristics.} How does the control plane behave under increasing load across three dimensions: (a) \emph{intent throughput}---requests per second at which decision latency degrades beyond the EQ1 threshold; (b) \emph{registry scale}---number of registered capabilities at which candidate generation becomes a bottleneck; and (c) \emph{policy set scale}---number of active policies at which evaluation cost dominates the decision cycle. Scaling curves along each dimension, measured independently and jointly, characterize the operational envelope of a Brain API deployment and identify where horizontal scaling, caching, or architectural changes are required.

Together, EQ1--EQ7 span five dimensions: correctness (EQ2), performance (EQ1, EQ4), optimality (EQ3), governability (EQ5), and scalability (EQ6, EQ7). A prototype evaluation addressing all seven would provide strong empirical grounding for the claims made in this paper and a complete characterization of the control plane's operational properties.

\subsection{Future Work: Agentic Datasets as a Specialization}
\label{sec:12.8}

\textbf{Impact.} Traditional data platforms treat datasets as passive artifacts that are queried by external agents but do not participate in decision-making about their own use. Agentic datasets invert this relationship: by registering capabilities, exposing context signals, and declaring governance policies, datasets become active participants in control-plane reasoning. This enables dynamic data-driven workflows that adapt to freshness and cost constraints without hard-coded routing logic, governance-aware data use in which compliance obligations are enforced before any query is issued, and autonomous analytical pipelines in which the control plane composes dataset nodes with model and tool nodes to fulfil complex intents end-to-end.

\textbf{Architecture positioning.} Brain API and agentic datasets are mutually reinforcing abstractions. Brain API provides the decision control plane---intent parsing, policy enforcement, capability selection, decision artifact production, and execution dispatch. Agentic datasets provide decision-aware data nodes: structured capability declarations, real-time context signals, and governance metadata that the control plane reasons over. The relationship is layered:

\begin{itemize}
\item \emph{Brain API} governs how and when dataset capabilities are invoked, without requiring datasets to embed control logic.
\item \emph{Agentic datasets} extend the capability registry with dataset-specific semantics (affordances, freshness, lineage), without requiring the control plane to understand storage internals.
\item \emph{Existing data infrastructure} is unchanged; the agentic dataset abstraction is a thin registration and context-exposure layer over existing backends.
\end{itemize}

This positions agentic datasets as a natural extension of the Brain API system model, not as a separate system. Viewed more broadly, Brain API can serve as a shared decision control plane for a research stack that also includes observability tooling (which monitors the decision plane), simulation environments (which exercise the policy engine under synthetic workloads), and domain-specific agents (which submit intents). Agentic datasets populate the capability registry for the data-centric tier of this stack.

\textbf{Concrete future directions.} A full treatment of agentic datasets would address: (a) a dataset descriptor schema (extending \texttt{capability\_spec} with \texttt{kind}, \texttt{affordances}, and lineage fields, as outlined in \S\ref{sec:10.1}); (b) an affordance matching algorithm that maps intent semantics to dataset affordances without requiring schema-level knowledge of each backend; (c) lineage-aware context providers that expose result freshness and provenance graphs as inputs to the decision engine; (d) governance-driven reuse policies that permit result reuse only when lineage and policy equivalence can be certified; and (e) empirical evaluation against the criteria defined in \S\ref{sec:12.7}, using a realistic dataset registry as the experimental substrate. We view this as the primary implementation and evaluation agenda following the foundational design presented here.

\section{Related Work}
\label{sec:13}

Brain API intersects multiple research and engineering domains, including distributed systems control planes, workflow and orchestration systems, policy engines, service meshes and API gateways, and agentic and tool-augmented AI systems. We briefly position our work relative to these areas and highlight the distinctions of an intent- and policy-aware control plane.

\subsection{Distributed Systems Control Planes}
\label{sec:13.1}

Systems such as Kubernetes, Mesos, and Borg introduced the separation of control plane and data plane for managing large-scale distributed resources \citep{burns2016kubernetes,hindman2011mesos,verma2015borg,schwarzkopf2013omega}. Kubernetes, in particular, provides declarative resource specifications and reconciliation loops for maintaining desired state. However, these systems focus primarily on \emph{resource state} (e.g., pods, services, deployments) rather than \emph{intent} or \emph{decision-making}. Brain API builds on the control-plane concept but elevates intent, policy, and capability selection to first-class concerns, targeting decision-level governance rather than only resource management.

\subsection{Workflow and Orchestration Engines}
\label{sec:13.2}

Workflow systems such as Airflow, Argo Workflows, Temporal, and Step Functions provide mechanisms for defining and executing multi-step computations \citep{deelman2005pegasus,argoworkflows,temporal}. These systems typically rely on statically defined graphs or imperative orchestration logic, with limited support for dynamic, policy-driven reconfiguration at runtime. While they excel at reliable execution and state management, the control logic that determines \emph{which} workflow or path to execute is usually external to the system. Brain API complements these engines by providing a decision layer that selects and synthesizes execution plans under explicit policy constraints, and by treating these decisions as auditable artifacts.

\subsection{Service Meshes and API Gateways}
\label{sec:13.3}

Service meshes and API gateways, such as Istio, Linkerd, and Envoy-based systems, provide traffic management, security, and observability at the network request level \citep{istio,linkerd,envoy}. Routing decisions are typically based on static rules, labels, or low-level metrics. While these systems offer powerful operational controls, they do not expose abstractions for reasoning over high-level intent or for governing tool and service selection based on semantic and policy considerations. Brain API operates at a higher semantic level, using intent and policy to drive routing and composition beyond individual network requests.

\subsection{Policy Engines}
\label{sec:13.4}

Policy engines such as Open Policy Agent (OPA) and Cedar provide declarative frameworks for expressing and evaluating authorization and governance rules \citep{opa,cedar2024,abadi1993logic,detreville2002binder}. These systems are widely used to enforce access control and compliance constraints. Brain API is complementary: it treats policy as a first-class input to control-plane decision-making, but it does not mandate a specific policy language or engine. Instead, it focuses on how policy evaluation is integrated into intent resolution, capability selection, and plan synthesis.

\subsection{Agentic and Tool-Augmented Systems}
\label{sec:13.5}

Recent agentic systems and tool-augmented frameworks integrate planners, large language models, and tool invocation to achieve high-level goals \citep{yao2023react,schick2023toolformer}. Examples include systems that perform tool routing, multi-step reasoning, and adaptive execution based on intermediate results. While these systems demonstrate the feasibility of intent-driven behavior, their control logic is often embedded in application code or framework-specific runtimes, and policy enforcement and observability are typically ad hoc. Brain API generalizes these patterns into an explicit control plane, separating decision-making from execution and providing governance and auditability as first-class features.

\subsubsection{Agent Orchestration Frameworks}
\label{sec:13.5.1}

Several frameworks have emerged that partially address the orchestration concern. \textbf{Microsoft Semantic Kernel} \citep{semantickernel} provides a plugin and planner model in which a kernel selects and composes functions to satisfy a goal. The planning step overlaps conceptually with Brain API's candidate generation and ranking stages. However, Semantic Kernel does not expose a declarative policy layer or produce auditable decision artifacts; governance logic must be embedded in plugins or host code. \textbf{LangGraph} \citep{langgraph} introduces explicit stateful control-flow graphs over LLM-based agents, enabling branching and looping that go beyond linear chains. Its graph edges model conditional transitions that can approximate policy-driven routing, but policies are expressed implicitly through branching logic rather than as first-class, independently auditable objects. \textbf{OpenAI Assistants API} \citep{openaiassistants} provides server-side tool routing and thread management, moving some orchestration logic out of client code. Yet policies, capability metadata, and routing rationale remain opaque to callers.

Brain API differs from all of these frameworks in that it exposes intent, policy, capability, and decision as independently addressable, first-class control-plane primitives. The decision process is observable and auditable independent of the planner or LLM implementation beneath it.

\subsection{Why Not Compose Existing Systems?}
\label{sec:13.6}

A recurring objection to a new abstraction is whether the same goals could be achieved by composing existing tools: OPA for policy evaluation, a service mesh for routing, a workflow engine for execution, and an LLM-based tool router for intent decomposition. We address this directly.

\textbf{The decision artifact does not exist in any component.} OPA evaluates a policy predicate and returns allow/deny; it does not produce a structured artifact that records the intent, the full policy set evaluated, context inputs used, and alternatives considered. A service mesh emits traces and metrics for requests that have already been dispatched---it cannot represent the reasoning that selected one capability over another before dispatch. Workflow engines record execution history, not decision history. Assembling these components produces operational observability but not \emph{decision-level} observability.

\textbf{Policy operates at the wrong layer in every component.} OPA governs individual authorization checks. Service mesh policies govern network-level traffic rules. Workflow engine conditions govern graph transitions. None of these enforce policy over the mapping from a high-level intent to a complete, multi-step execution plan---the level at which agentic governance is needed. Brain API's policy layer is evaluated once per intent, before any action is dispatched, and its constraints propagate through plan synthesis. This is architecturally distinct from per-request or per-transition policy evaluation.

\textbf{No shared semantic for capability selection exists.} In a composed stack, the workflow engine, tool router, and service mesh each have their own models for what a "capability" is and how to select one. Brain API's capability registry provides a unified, semantically annotated abstraction over all of these, enabling intent-driven selection across heterogeneous backends under a single policy and decision model.

\textbf{Integration creates fragmentation, not unification.} Each additional component introduces its own observability model, API surface, and failure semantics. Correlating a decision rationale across OPA logs, service mesh traces, and workflow execution records requires bespoke integration work that must be repeated for each deployment. Brain API provides this correlation as a first-class architectural property through stable identifiers that span intent, decision, plan, and execution.

\subsection{Semantic Routing and Caching}
\label{sec:13.7}

Work on semantic routing, vector-based retrieval, and result caching explores how similarity metrics and learned representations can accelerate or adapt execution \citep{lewis2020rag,khandelwal2020knnlm}. Such techniques are increasingly used in AI-assisted systems to reuse prior results or select tools based on semantic compatibility. Brain API does not prescribe a particular semantic mechanism, but it provides a control-plane context in which semantic signals can influence routing and planning under policy constraints. We view dedicated semantic memory or caching planes as complementary extensions to the core control-plane model presented here.

\subsection{Comparative Summary}
\label{sec:13.8}

Table~\ref{tab:comparison} contrasts Brain API against the systems surveyed above across six dimensions that are central to the problem Brain API addresses. A checkmark ($\checkmark$) indicates native support; a partial mark ($\circ$) indicates partial or ad hoc support that requires application-level code; and a dash (--) indicates the concern is out of scope for the system.

\begin{table}[ht]
\centering
\caption{Comparison of Brain API against related systems across six control-plane dimensions.
$\checkmark$ = native support; $\circ$ = partial/ad hoc support; -- = out of scope.}
\label{tab:comparison}
\resizebox{\textwidth}{!}{%
\begin{tabular}{lcccccc}
\toprule
\textbf{System} &
\textbf{\shortstack{Intent as\\primitive}} &
\textbf{\shortstack{First-class\\decl.\ policy}} &
\textbf{\shortstack{Auditable\\decision artifacts}} &
\textbf{\shortstack{Decision-level\\telemetry}} &
\textbf{\shortstack{Heterogeneous\\backends}} &
\textbf{\shortstack{Semantic\\capability routing}} \\
\midrule
Kubernetes / Borg / Mesos          & --            & $\circ$       & --            & --            & $\checkmark$  & --            \\
Workflow engines (Airflow, Argo, Temporal) & --     & $\circ$       & $\circ$       & $\circ$       & $\checkmark$  & --            \\
Service meshes / API gateways (Istio, Envoy) & --  & $\circ$       & --            & $\checkmark$  & $\circ$       & --            \\
Policy engines (OPA, Cedar)        & --            & $\checkmark$  & $\circ$       & --            & --            & --            \\
Semantic Kernel                    & $\circ$       & --            & --            & --            & $\circ$       & $\circ$       \\
LangGraph                          & $\circ$       & --            & --            & --            & $\circ$       & $\circ$       \\
OpenAI Assistants API              & $\circ$       & --            & --            & --            & --            & $\circ$       \\
\midrule
\textbf{Brain API (proposed)}      & $\checkmark$  & $\checkmark$  & $\checkmark$  & $\checkmark$  & $\checkmark$  & $\checkmark$  \\
\bottomrule
\end{tabular}%
}
\end{table}

The table shows that individual systems address at most two or three of these dimensions natively. Systems in the distributed infrastructure column (Kubernetes, service meshes) handle heterogeneous backends and some observability but lack intent and semantic routing. Agentic frameworks (Semantic Kernel, LangGraph) provide partial intent and routing support but have no declarative policy primitives or auditable decision artifacts. Policy engines provide strong policy evaluation but are silent on the remaining dimensions. Brain API is distinguished by addressing all six dimensions within a single, unified control-plane abstraction---a combination that does not exist in any currently deployed system surveyed here.

\section{Conclusion and Future Work}
\label{sec:14}

In this paper, we proposed and argued for Brain API, an intent-aware control plane for policy-governed agentic systems. The central contribution is the introduction of \emph{decision artifacts} as a new primitive in distributed systems control planes. Traditional control planes manage resources, tasks, deployments, and services; Brain API manages \emph{decisions}---durable, auditable objects that record the complete reasoning process used to transform a high-level intent into an executable plan, including which policies applied, which capabilities were evaluated, and why a particular execution path was chosen over its alternatives. By elevating decisions to first-class control-plane objects, Brain API enables a class of governance capabilities that existing systems cannot provide: post-hoc compliance auditing against specific policy versions, reproducible reasoning that can be replayed under evolved policies, forensic debugging of unexpected system behavior, and intent-level governance that enforces constraints before any action is taken.

We detailed the core control-plane interfaces, the decision and routing semantics, and the execution model that together form the proposed architecture for governed, adaptive, and observable behavior across heterogeneous backends. Through representative scenarios, we illustrated how Brain API simplifies the construction and governance of agentic systems that would otherwise require ad hoc orchestration logic and fragmented policy enforcement.

We then evaluated a prototype of the decision layer against two policy corpora we did not author, the OPA Gatekeeper constraint library and Cedar's published example policies, whose expected outcomes come from a third party: the library's own published verdicts and Cedar's reference implementation (\S\ref{sec:11}). Two findings deserve restating. The first is negative and concerns our own design: a rule language of scalar comparisons expressed 3 of 49 real policies, because governance policies quantify over collections, and a subsequent conformance failure revealed a second deficiency---the absence of per-element conjunction---that no amount of inspection had surfaced. Both were corrected, and both were found only because the corpus was not ours. The second is that decision artifacts and admission logs differ in what an auditor can recover from them: six of six audit questions against two, with every decision re-derivable from its artifact alone. The questions an admission log cannot answer are exactly those about alternatives, which is a structural consequence of adjudicating one proposal rather than selecting among several.

Several directions for future work follow. First, richer semantic memory and result-reuse mechanisms can be integrated as first-class context providers, enabling more sophisticated forms of reuse, caching, and learning from prior executions under explicit policy control. Second, more formal policy languages and verification techniques could be applied to reason about the correctness, safety, and completeness of policy sets and decision procedures; the aggregation and regular-expression constructs currently outside our rule grammar are a concrete starting point. Third, the evaluation here covers policy filtering and selection in two policy-adjudication domains; candidate generation, context signals, ranking and plan synthesis are not measured, and quantifying decision latency and policy-evaluation cost under production load is the most immediate next step.

More broadly, we believe that elevating decisions to durable control-plane objects opens a path toward distributed agentic systems that are not only scalable and reliable, but also governable, inspectable, and auditable by design. Brain API represents a step in this direction: a unifying decision-centric control plane in which intent, policy, and reasoning are first-class concerns, and in which every execution path is traceable to the decision that produced it.

\bibliographystyle{unsrtnat}
\bibliography{references}

\end{document}